\documentclass[a4paper,12pt]{article}
\usepackage[a4paper,text={16.8cm,22.4cm}]{geometry}
\usepackage{amsmath,amssymb,bm,psfrag,graphicx}
\usepackage[normalem]{ulem}

\allowdisplaybreaks 
\begin{document}

\begin{titlepage}

\begin{flushright}
MZ-TH/11-29\\
Draft \today
\end{flushright}

\vspace{0.2cm}
\begin{center}
\Large\bf
Electroweak Gauge-Boson Production at Small $\bm{q_T}$:\\
Infrared Safety from the Collinear Anomaly
\end{center}

\vspace{0.2cm}
\begin{center}
Thomas Becher$^a$, Matthias Neubert$^b$ and Daniel Wilhelm$^b$\\
\vspace{0.4cm}
{\sl 
${}^a$\,Institut f\"ur Theoretische Physik, Universit\"at Bern\\
Sidlerstrasse 5, CH--3012 Bern, Switzerland\\[0.3cm]
${}^b$\,Institut f\"ur Physik (THEP), 
Johannes Gutenberg-Universit\"at\\ 
D--55099 Mainz, Germany}
\end{center}

\vspace{0.2cm}
\begin{abstract}
\vspace{0.2cm}
\noindent 
Using methods from effective field theory, we develop a novel, systematic framework for the calculation of the cross sections for electroweak gauge-boson production at small and very small transverse momentum $q_T$, in which large logarithms of the scale ratio $M_V/q_T$ are resummed to all orders. These cross sections receive logarithmically enhanced corrections from two sources: the running of the hard matching coefficient and the collinear factorization anomaly. The anomaly leads to the dynamical generation of a non-perturbative scale $q_*\sim M_V\,e^{-{\rm const}/\alpha_s(M_V)}$, which protects the processes from receiving large long-distance hadronic contributions. Expanding the cross sections in either $\alpha_s$ or $q_T$ generates strongly divergent series, which must be resummed. As a by-product, we obtain an explicit non-perturbative expression for the intercept of the cross sections at $q_T=0$, including the normalization and first-order $\alpha_s(q_*)$ correction. We perform a detailed numerical comparison of our predictions with the available data on the transverse-momentum distribution in $Z$-boson production at the Tevatron and LHC. 
\end{abstract}
\vfil

\end{titlepage}

\section{Introduction}

In collider processes with several disparate scales, fixed-order perturbative expansions in QCD become unreliable since higher-order corrections are enhanced by large double logarithms of scale ratios. The classic example of such a multi-scale process is the Drell-Yan production of electroweak gauge bosons with transverse momentum $q_T$ much smaller than their mass. The leading logarithmically-enhanced corrections in this kinematic region were resummed in \cite{DDT,Parisi:1979se,Curci:1979bg}, and an all-order formula for the resummed cross section at small $q_T$ was obtained in the seminal work \cite{Collins:1984kg}. The region of small $q_T$ is of great phenomenological importance, since it has the largest cross section and is used e.g.\ to extract the $W$-boson mass and width. In the related process of Higgs-boson production via gluon fusion, the region of small $q_T$ is important because one usually vetoes hard jets in order to enhance the signal over background ratio. The traditional resummation approach of \cite{Collins:1984kg} suffers from the presence of singularities arising from integrals over the Landau pole of the running coupling constant, and hence a prescription is required to regularize the integral. In practical applications, the integration is cut off at large $x_T$ values, and to account for the missing contributions a non-perturbative model function is employed. For transverse momenta in the perturbative domain these long-distance contributions are formally power suppressed, but it is irritating that an explicit prescription for how to deal with them is needed even for $q_T$ values deep in the perturbative regime. The explicit cut-off also makes it difficult to perform the matching onto fixed-order computations, since cut-off effects persist even when the formula is evaluated at large $q_T\sim M_V$, where $M_V$ denotes the mass of the Drell-Yan object.

In a recent paper, we have derived a novel form of the all-order factorization theorem for Drell-Yan production at small transverse momentum, $\Lambda_{\rm QCD}\ll q_T\ll M_V$ \cite{Becher:2010tm}, using methods from effective field theory. The cross section is expressed as a product of a $q^2$-dependent hard function with a convolution of two transverse-position ($x_T$) dependent parton distribution functions (PDFs). The form of the factorization theorem is affected by an anomaly of the effective Lagrangian, which implies that the naive factorization of two collinear sectors valid at the level of the classical Lagrangian is broken by quantum effects. This gives rise to an anomalous, $q^2$-dependent factor under the convolution integral in $x_T$ space. Consistency conditions ensure that this factor is a pure power of $q^2$, with an $x_T$-dependent exponent. In the short-distance region, for $x_T\ll\Lambda_{\rm QCD}^{-1}$, the product of transverse-position dependent PDFs can be expanded in terms of standard PDFs convoluted with perturbatively calculable kernel functions. The resulting effective field-theory formula for the resummed cross section is free of Landau-pole singularities and therefore {\em per se\/} does not require non-perturbative modeling. 

Naively, assuming that transverse momentum $q_T$ and transverse displacement $x_T$ are conjugate variables satisfying $q_T x_T\sim 1$, the factorization formula derived in \cite{Collins:1984kg} and \cite{Becher:2010tm} only applies as long as $q_T\gg\Lambda_{\rm QCD}$ is in the perturbative domain. However, as early as in 1979, Parisi and Petronzio have argued that for asymptotically large momentum transfer, the Drell-Yan cross section at very small and even vanishing transverse momentum of the lepton pair can be calculated in resummed perturbation theory \cite{Parisi:1979se}. Using methods from effective field theory, we develop a novel framework for the systematic short-distance calculation of the Drell-Yan cross section at small and very small $q_T$. We show that if the mass $M_V$ of the Drell-Yan object is so large that the scale $q_*\approx M_V\exp\left[-\frac{2\pi}{(4C_R+\beta_0)\,\alpha_s(M_V)}\right]$ is in the perturbative domain (here $R=F,A$ denotes the color representation of the partons creating the object), the cross section can be calculated using renormalization-group (RG) improved perturbation theory for arbitrarily small values of $q_T$, up to power corrections controlled by $\Lambda_{\rm QCD}/q_*$. The scale $q_*$ emerges dynamically, and it screens the cross section from receiving long-distance contributions at leading power. We distinguish three regions of transverse momentum, which require different classes of terms to be resummed. In the region of large $q_T$, approximately $q_T>15$\,GeV for the case of electroweak gauge-boson production, a fixed-order perturbative calculation is justified. In an intermediate region, the perturbative series exhibits large logarithms as well as a strong factorial growth of a certain set of higher-order corrections, and both must be resummed in order to obtain reliable results. Finally, in the region of very small $q_T\lesssim q_*$, the power counting of the perturbative expansion must be modified and certain higher-order terms must be resummed at all orders, even though they are naively suppressed by powers of $\alpha_s$. We set up our formalism in such a way that a single formula  automatically interpolates between the three regions. We provide all ingredients to perform the resummation of large logarithms at next-to-next-to-leading logarithmic (NNLL) order. In the low-$q_T$ region this requires retaining terms of up to four-loop order in some coefficient functions. These terms can be related to known anomalous dimensions using RG equations.

As a by-product of our analysis, we derive for the first time an explicit formula for the intercept of the differential cross section $d\sigma/dq_T^2$ at vanishing transverse momentum, in which the normalization and first-order perturbative correction in $\alpha_s(q_*)$ are included. We show that in the region of very small $q_T$ a usual operator-product expansion (OPE) of the cross section is inapplicable due to a very strong asymptotic divergence of the twist expansion. The dependence on $q_T$ and the sensitivity to long-distance hadronic effects can only be assessed by resumming the OPE. For the case of the $q_T$ dependence such a resummation is implicit in our approach. Long-distance corrections, which could affect the matching of transverse-position dependent PDFs onto standard PDFs, can at present only be modeled using a phenomenological function $f_{\rm hadr}(x_T\Lambda_{\rm NP})$. Our formalism provides a convenient framework for implementing such a form factor. Using different forms of model functions, we find that to a good accuracy the shape of the cross section is only affected by the coefficient of the first term in the expansion of the form factor in transverse separation, $f_{\rm hadr}(x_T\Lambda_{\rm NP})=1-\Lambda_{\rm NP}^2\,x_T^2+\dots$. The main effect of long-distance corrections is to shift the peak of the $d\sigma/dq_T$ distribution for $Z$ bosons produced at the Tevatron and LHC by roughly about $\Lambda_{\rm NP}$. We do not confirm the rather pessimistic statements about the validity of the short-distance analysis of the transverse-momentum distribution in the region of very small $q_T$ made in \cite{Collins:1984kg}, where it was concluded that a Drell-Yan mass as large as $10^8$\,GeV would be required to keep power corrections to the intercept at an acceptable level (below 20\%). Instead, we find that the theory developed in this paper works well for $W$ and $Z$ production at hadron colliders, and it will work even better for the production of Higgs bosons via gluon fusion or the production of new heavy particles such as $W'$ or $Z'$ bosons or slepton pairs.

In Section~\ref{sec:fact} we briefly review results from our previous work \cite{Becher:2010tm} and reorganize them in a way suited for our discussion. In Section~\ref{sec:expansion} we then explain the systematics of the resummation scheme valid at very small $q_T$, where the emergent non-perturbative short-distance scale $q_*$ protects the cross section from receiving long-distance contributions at leading power. The modified power counting required in this region is introduced and used to derive the relevant expansions of the various coefficient functions in the factorization formula for the differential cross section. We also explain the relevance of the three regions in transverse-momentum space mentioned above. Section~\ref{sec:intercept} is devoted to a study of the $q_T$ distribution near the origin. We derive an explicit formula for the cross section $d\sigma/dq_T^2$ at $q_T=0$ including the normalization and first-order correction in $\alpha_s(q_*)$. We also explain why the OPE breaks down for very small $q_T$. Detailed systematics studies of of our results are performed in Section~\ref{sec:numerics}, while Section~\ref{sec:data} is devoted to comparisons with Tevatron and LHC data on $Z$-boson production. We conclude in Section~\ref{sec:concl}. Technical details of our calculations are described in four appendices.

\section{Factorization and resummation}
\label{sec:fact}

In a recent paper \cite{Becher:2010tm}, we have analyzed the Drell-Yan process using methods of effective field theory. In the kinematical region where $\Lambda_{\rm QCD}\ll q_T\ll M_Z$, the double differential cross section for $Z$-boson production was shown to obey the factorization formula
\begin{equation}\label{fact1}
\begin{aligned}
   \frac{d^2\sigma}{dq_T^2\,dy} 
   &= \frac{4\pi^2\alpha}{N_c\,s} \left| C_V(-M_Z^2,\mu) \right|^2
    \sum_q\,\frac{|g_L^q|^2+|g_R^q|^2}{2}\,\sum_{i=q,g} \sum_{j=\bar q,g} 
    \int_{\xi_1}^1\!\frac{dz_1}{z_1} \int_{\xi_2}^1\!\frac{dz_2}{z_2} \\
   &\quad\times \bigg[ \bar C_{q\bar q\leftarrow ij}(z_1,z_2,q_T^2,M_Z^2,\mu)\,
    \phi_{i/N_1}(\xi_1/z_1,\mu)\,\phi_{j/N_2}(\xi_2/z_2,\mu) 
    + (q,i\leftrightarrow\bar q,j) \bigg] \,,
\end{aligned}
\end{equation}
where $g_{L,R}^q$ denote the $Z$-boson couplings to quarks $q_{L,R}$ in units of $e$, $\phi_{i/N}(z,\mu)$ are standard PDFs, and $\xi_{1,2}=\sqrt{\tau}\,e^{\pm y}$, where $\tau=(M_Z^2+q_T^2)/s$ and $y$ is the rapidity of the $Z$ boson in the laboratory frame. A corresponding formula for the differential cross section $d\sigma/dq_T^2$, along with explicit expressions for the weak charges, is presented in Appendix~\ref{app:a}. The function $C_V$ is a short-distance (``hard'') Wilson coefficient arising in the matching of the weak currents of the $Z$ boson onto the leading-power current operator in soft-collinear effective theory (SCET) \cite{Manohar:2003vb,Becher:2006nr}. The perturbative (``collinear'') kernel functions $\bar C_{q\bar q\leftarrow ij}$ are given by (we denote $q_T^2\equiv-q_\perp^2$ and $x_T^2\equiv-x_\perp^2$)
\begin{equation}\label{Cdef}
\begin{aligned}
   \bar C_{q\bar q\leftarrow ij}(z_1,z_2,q_T^2,M_Z^2,\mu)
   &= \frac{1}{4\pi} \int\!d^2x_\perp\,e^{-iq_\perp\cdot x_\perp}
    \left( \frac{x_T^2 M_Z^2}{b_0^2} \right)^{-F_{q\bar q}(L_\perp,a_s)} \\
   &\quad\times I_{q\leftarrow i}(z_1,L_\perp,a_s)\,
    I_{\bar q\leftarrow j}(z_2,L_\perp,a_s) \,,
\end{aligned}
\end{equation}
where 
\begin{equation}\label{defs}
   a_s = \frac{\alpha_s(\mu)}{4\pi} \,, \qquad
   L_\perp = \ln\frac{x_T^2\mu^2}{b_0^2} \,, \qquad b_0 = 2e^{-\gamma_E} \,.
\end{equation}
The functions $I_{i\leftarrow j}$ arise in the matching of transverse-position dependent PDFs $B_{i/N}$ defined in \cite{Becher:2010tm} onto ordinary PDFs, 
\begin{equation}\label{OPE}
   B_{i/N}(\xi,x_T^2,\mu) 
   = \sum_j \int_\xi^1\!\frac{dz}{z}\,I_{i\leftarrow j}(z,x_T^2,\mu)\,
    \phi_{j/N}(\xi/z,\mu) + {\cal O}(\Lambda_{\rm QCD}^2\,x_T^2) \,, 
\end{equation}
which is valid at small transverse separation $x_T^2\ll\Lambda_{\rm QCD}^2$. The factorized cross section (\ref{fact1}) receives power corrections in the two small ratios $q_T^2/M_Z^2$ and $\Lambda_{\rm QCD}^2/q_T^2$, where the latter ones enter via (\ref{OPE}). These power corrections will not be indicated explicitly in most of our equations. 

In the factorization formulas (\ref{fact1}) and (\ref{Cdef}), the dependence on the two disparate scales $M_Z$ and $q_T$ is factorized explicitly in $x_T$ space. Note the unusual fact that dependence on the high scale $M_Z$ enters in two places: via the hard matching coefficient $C_V$, but also via an $x_T$-dependent power of $M_Z$ under the Fourier integral in (\ref{Cdef}). The latter effect is due to the collinear factorization anomaly discovered in \cite{Becher:2010tm} (see also \cite{Beneke_talk}), which results from the fact that the naive factorization property of the classical effective Lagrangian of SCET (i.e., the property that different sectors of the effective theory do not interact with one another) is in some cases spoiled by quantum effects. This anomaly introduces a power-law dependence on $M_Z$ in the kernel functions in $x_T$ space. The anomalous exponent $F_{q\bar q}$ is constrained by the non-abelian exponentiation theorem \cite{Gatheral:1983cz,Frenkel:1984pz}. As long as $x_T^2\ll\Lambda_{\rm QCD}^2$, it can be calculated in perturbation theory, and it is presently known to two-loop order \cite{Becher:2010tm}. A simpler example for the occurrence of the collinear anomaly is the Sudakov form factor of a massive vector boson, for which the anomalous power-law dependence on the gauge-boson mass was derived in \cite{Chiu:2007dg}. More recently, it was shown that also the jet-broadening distribution in $e^+ e^-$ annihilations is affected by a factorization anomaly \cite{Becher:2011pf}.

The factorization formulas (\ref{fact1}) and (\ref{Cdef}) have been shown in \cite{Becher:2010tm} to be equivalent, to all orders in perturbation theory, to an expression for the resummed cross section derived in a seminal work by Collins, Soper, and Sterman (CSS) \cite{Collins:1984kg}. An interesting effect of the anomalous terms is that they give rise to additional contributions to the resummation exponents $A$ and $B$ in the CSS approach, which therefore do not agree with the corresponding quantities for soft-gluon resummation. In particular, starting at three-loop order the coefficient $A$ is longer equal to the cusp anomalous dimension $\Gamma_{\rm cusp}^F$. We note at this point that recently an alternative SCET-based resummation scheme for Drell-Yan production at small $q_T$ was presented \cite{Mantry:2009qz,Mantry:2010mk}, in which only the large logarithms contained in the hard function $|C_V|^2$ in (\ref{fact1}) are resummed. This approach does not address the resummation of the large logarithms steming from the collinear anomaly, which reside in the kernel functions $\bar C_{q\bar q\leftarrow ij}$ in (\ref{Cdef}). The method of \cite{Mantry:2009qz,Mantry:2010mk} therefore does not provide a consistent resummation scheme already at NLL order. While these authors correctly reproduce the next-to-leading logarithms in the cross section itself, this is not sufficient. For observables subject to Sudakov double logarithms, it is essential that the counting of logarithms is performed in the exponent and not at the level of the cross section. If the NLL terms are not exponentiated, higher-order terms in perturbation theory become arbitrarily large in the region where the logarithms are of ${\cal O}(1/\alpha_s)$. In Appendix~\ref{nloexpansion} we discuss the systematics of the expansion in more detail and give explicit expressions for the fixed-order expansion to ${\cal O}(\alpha_s^2)$ of our result for the cross section. We also show explicitly which of these terms contain the large logarithms {\rm not\/} accounted for by the resummation of the hard function.

The resummation of large logarithms in the factorization formula (\ref{fact1}) is accomplished by evolving the hard matching coefficient $C_V$ to a scale determined by the {\em average\/} transverse separation, $\mu\sim\langle x_T^{-1}\rangle\ll M_Z$, in a sense described in more detail in the next section.  With this scale choice the logarithm $L_\perp$ in (\ref{defs}) is of ${\cal O}(1)$, and hence the functions $F_{q\bar q}$ and $I_{q\leftarrow i}$ can be calculated using fixed-order perturbation theory. All large logarithms are then contained in the coefficient $C_V(-M_Z^2,\mu)$ and in the anomalous factor $(x_T^2 M_Z^2)^{-F_{q\bar q}}$. The RG evolution equation of the hard matching coefficient at time-like momentum transfer $q^2$ is of the Sudakov type and reads \cite{Becher:2006nr}
\begin{equation}\label{Cevol}
   \frac{d}{d\ln\mu}\,C_V(-q^2,\mu) 
   = \left[ \Gamma_{\rm cusp}^F(a_s)\,\ln\frac{-q^2}{\mu^2} + 2\gamma^q(a_s) \right] 
    C_V(-q^2,\mu) \,,
\end{equation}
where $\Gamma_{\rm cusp}^F$ is the cusp anomalous dimension in the fundamental representation and $\gamma^q$ denotes the anomalous dimension of a collinear quark field in SCET. These quantities are known to three-loop order. The explicit form of the solution to this equation up to next-to-next-to-leading order (NNLO) in RG-improved perturbation theory has been discussed in detail in \cite{Becher:2006mr}. The advantages of using a time-like scale choice ($\mu_h^2<0$) for time-like processes such as Drell-Yan production were emphasized in \cite{Ahrens:2008qu,Ahrens:2008nc,Magnea:1990zb}. In Appendix~\ref{app:hard}, we compile the NLO expression for the hard matching coefficient needed for our analysis. RG invariance of the cross section (\ref{fact1}) requires that the cusp logarithm $2\Gamma_{\rm cusp}^F\ln M_Z^2$ resulting from the scale variation of the hard function be compensated by a corresponding term in the scale variation of the kernels $\bar C_{q\bar q\leftarrow ij}$. This is ensured by the RG equation \cite{Becher:2010tm}
\begin{equation}\label{Fevol}
   \frac{d}{d\ln\mu}\,F_{q\bar q}(L_\perp,a_s) = 2\Gamma_{\rm cusp}^F(a_s)
\end{equation}
for the anomalous exponent. Even though it is not required for the resummation procedure, it will be important to also consider the evolution equations for the kernel functions $I_{q\leftarrow i}$. They are given by 
\begin{equation}
\begin{aligned}
   \frac{d}{d\ln\mu}\,I_{q\leftarrow i}(z,L_\perp,a_s)
   &= \Big[ \Gamma_{\rm cusp}^F(a_s)\,L_\perp - 2\gamma^q(a_s) \Big]\,
    I_{q\leftarrow i}(z,L_\perp,a_s) \\
   &\quad\mbox{}- \sum_j \int_z^1\!\frac{du}{u}\,I_{q\leftarrow j}(u,L_\perp,a_s)\, 
    {\cal P}_{j\leftarrow i}(z/u,a_s) \,,
\end{aligned}
\end{equation}
where ${\cal P}_{j\leftarrow i}$ are the usual DGLAP splitting functions. The first term on the right-hand side implies that the functions $I_{q\leftarrow i}$ exhibit double logarithmic dependence on $L_\perp$ in the exponent. It will be important for our purposes to factor out these terms, and this can be accomplished by rewriting
\begin{equation}
   I_{q\leftarrow i}(z,L_\perp,a_s)
   \equiv e^{h_F(L_\perp,a_s)}\,\bar I_{q\leftarrow i}(z,L_\perp,a_s) \,,
\end{equation}
where
\begin{equation}\label{hevol}
   \frac{d}{d\ln\mu}\,h_F(L_\perp,a_s)
   = \Gamma_{\rm cusp}^F(a_s)\,L_\perp - 2\gamma^q(a_s) \,.
\end{equation}
We choose to define $h_F(0,a_s)\equiv 0$, so that $h_F(L_\perp,a_s)$ contains logarithms of $L_\perp$ only. The new functions $\bar I_{q\leftarrow i}$ now evolve exactly like the usual PDFs (but with the opposite sign in front of the DGLAP splitting functions), while $h_F$ contains all double-logarithmic terms. We can now rewrite the hard-scattering kernels from (\ref{Cdef}) in the form
\begin{equation}\label{Cfinal}
\begin{aligned}
   \bar C_{q\bar q\leftarrow ij}(z_1,z_2,q_T^2,M_Z^2,\mu)
   &= \frac12 \int_0^\infty\!dx_T\,x_T\,J_0(x_T q_T)\,
    \exp\Big[ g_F(M_Z^2,\mu,L_\perp,a_s) \Big] \\
   &\quad\times \bar I_{q\leftarrow i}(z_1,L_\perp,a_s)\,
    \bar I_{\bar q\leftarrow j}(z_2,L_\perp,a_s) \,, \\
\end{aligned}
\end{equation}
where
\begin{equation}\label{gdef}
   g_F(M_Z^2,\mu,L_\perp,a_s) 
   = - \left( \ln\frac{M_Z^2}{\mu^2} + L_\perp \right) F_{q\bar q}(L_\perp,a_s)
    + 2 h_F(L_\perp,a_s) \,.
\end{equation}

As mentioned above, our strategy will be to choose the factorization scale $\mu\sim\langle x_T^{-1}\rangle\ll M_Z$, in such a way that the functions $F_{q\bar q}$ and $h_F$, as well as the kernels $\bar I_{q\leftarrow i}$, can be evaluated in fixed-order perturbation theory. When using the resulting expressions to write down the perturbative expansion of the exponent $g_F$, we treat $\ln(M_Z^2/\mu^2)$ as a large logarithm and count
\begin{equation}\label{etadef}
   \eta\equiv \eta_F(M_Z^2,\mu) = \Gamma_0^F a_s \ln\frac{M_Z^2}{\mu^2}
   = \frac{C_F\alpha_s(\mu)}{\pi}\,\ln\frac{M_Z^2}{\mu^2}
\end{equation}
as an ${\cal O}(1)$ variable. Here $\Gamma_0^F=4C_F$ is the one-loop coefficient of the cusp anomalous dimension in the fundamental representation of SU$(N_c)$. We then obtain
\begin{equation}\label{gFexp}
\begin{aligned}
   g_F(\eta,L_\perp,a_s) 
   &= - \eta L_\perp - a_s \left[ \left( \Gamma_0^F + \eta\beta_0 \right) \frac{L_\perp^2}{2}   
    + \left( 2\gamma_0^q + \eta K \right) L_\perp + \eta d_2 \right]
    + {\cal O}(a_s^2) \,,
\end{aligned}
\end{equation}
where $\gamma_0^q=-3C_F$ and $\beta_0=\frac{11}{3}\,C_A-\frac43\,T_F n_f$ are the one-loop coefficients of the quark anomalous dimension and $\beta$-function, and with a slight abuse of notation we have eliminated the arguments $M_Z^2$ and $\mu$ in $g_F$ in favor of $\eta$. The quantities
\begin{equation}
   K = \frac{\Gamma_1^F}{\Gamma_0^F} 
    = \left( \frac{67}{9} - \frac{\pi^2}{3} \right) C_A - \frac{20}{9}\,T_F n_f \,, \qquad
   d_2 = \frac{d_2^q}{\Gamma_0^F} 
    = \left( \frac{202}{27} - 7\zeta_3 \right) C_A - \frac{56}{27}\,T_F n_f
\end{equation}
contain some two-loop information. These particular ratios are the same for any representation of the gauge group. Finally, the one-loop expressions for the kernel functions $\bar I_{q\leftarrow i}$ in (\ref{Cfinal}) read
\begin{equation}\label{barIexp}
   \bar I_{q\leftarrow i}(z,L_\perp,a_s) 
   = \delta(1-z)\,\delta_{qi} - a_s \left[ {\cal P}_{q\leftarrow i}^{(1)}(z)\,\frac{L_\perp}{2} 
    - {\cal R}_{q\leftarrow i}(z) \right] + {\cal O}(a_s^2) \,,
\end{equation}
where
\begin{equation}\label{APkernels}
   {\cal P}_{q\leftarrow q}^{(1)}(z) 
   = 4C_F \left( \frac{1+z^2}{1-z} \right)_+ , \qquad
   {\cal P}_{q\leftarrow g}^{(1)}(z) 
   = 4T_F \left[ z^2 + (1-z)^2 \right]
\end{equation}
are the one-loop DGLAP splitting functions, and the remainder functions
\begin{equation}
   {\cal R}_{q\leftarrow q}(z) 
   = C_F \left[ 2(1-z) - \frac{\pi^2}{6}\,\delta(1-z) \right] , \qquad
   {\cal R}_{q\leftarrow g}(z) 
   = 4T_F\,z(1-z)
\end{equation}
can be extracted from results obtained in \cite{Becher:2010tm}.

\section{Scale setting and systematics of the expansion}
\label{sec:expansion}

A reliable evaluation of the factorization formula (\ref{fact1}) requires that the factorization scale $\mu$ be chosen such that the logarithm $L_\perp=\ln(x_T^2\mu^2/b_0^2)$ entering the various coefficient functions in (\ref{Cfinal}) is a small quantity. The most naive choice would be to set $\mu\sim b_0/x_T$ inside the integral over the Bessel function, so that $L_\perp$ is a small logarithm for any choice of $x_T$. There are several disadvantages to that treatment. First, since $x_T$ is integrated over all possible values, there is no clear meaning to the scale $\mu$ in the sense of a physical, characteristic scale of the process. Second, setting the scale inside the integral implies that the integration unavoidably hits the Landau pole of the running coupling, giving rise to ambiguities in the numerical results. In the spirit of effective field theory, the scale $\mu$ should correspond to a physical scale in the underlying factorization theorem. In our case, where we have chosen to evolve the hard function down to a typical collinear scale, the requirement is that $\mu$ should be set such that the perturbative series for the kernels $\bar C_{q\bar q\leftarrow ij}$ has a well-behaved perturbative expansion. Obviously, this requires that {\em on average\/} the $x_T$-dependent logarithm $L_\perp$ is small, so that the perturbative expansions of the functions $g_F$ and $\bar I_{q\leftarrow i}$ are well behaved.

The factorization formula (\ref{fact1}) has been derived under the assumption $\Lambda_{\rm QCD}\ll q_T\ll M_Z$. It neglects power corrections of order $q_T^2/M_Z^2$, and also higher-order terms $\sim\Lambda_{\rm QCD}^2\,x_T^2$ in (\ref{OPE}). The first type of power corrections can be included by matching our formulas to the known fixed-order results for the differential cross section (see below). Naively, one would expect that the transverse momentum $q_T$ and transverse separation $x_T$ are conjugate variables satisfying $q_T x_T\sim 1$, in which case the second class of power corrections would scale like $\Lambda_{\rm QCD}^2/q_T^2$. While this is sometimes true, the general situation turns out to be more complicated. After integration over $x_\perp$, the factorized dependence on $M_Z$ and $q_T$ in (\ref{Cdef}) gets intertwined in a complicated way, and this gives rise to the peculiar effect that the two scales $q_T$ and $x_T$ decouple for very small $q_T$ as long as $M_Z$ is sufficiently large. This explains an observation made long ago by Parisi and Petronzio, who found that the intercept of the differential Drell-Yan cross section $d\sigma/dq_T^2$ at $q_T=0$ can be calculated using short-distance methods provided that the mass $M_V$ of the Drell-Yan pair is asymptotically large \cite{Parisi:1979se}. Note that without the collinear anomaly (i.e., for $F_{q\bar q}=0$) such an effect could not arise, since then the scales $q_T$ and $x_T$ would trivially be related by $q_\perp\cdot x_\perp\sim 1$ due to the Fourier integral in (\ref{Cdef}). The interesting interplay beteween the scales $M_V$ and $q_T$ is less obvious in the traditional formalism, which implicitly adopts the choice $\mu\sim b_0/x_T$. This choice eliminates the anomaly at NLL, but the $M_V$ dependence then enters via the hard function, which for $\mu\sim b_0/x_T$ is a non-trivial part of the Fourier integral.

To analyze the Fourier integral in the effective theory, consider the leading-order approximation for the kernels $\bar C_{q\bar q\leftarrow ij}$ in (\ref{Cfinal}), where we set $a_s\to 0$ and only keep the first term in the perturbative series for the exponent $g_F$ in (\ref{gFexp}). This gives rise to the integral
\begin{equation}\label{toy}
   \frac12 \int_0^\infty\!dx_T\,x_T\,J_0(x_T q_T)\,e^{-\eta L_\perp}
   =\frac{1}{q_T^2} \left( \frac{q_T^2}{\mu^2} \right)^\eta
    \frac{\Gamma(1-\eta)}{e^{2\eta\gamma_E}\,\Gamma(\eta)} \,.
\end{equation}
The integral over the Bessel function converges at large distances ($x_T\to\infty$) only if $\eta>\frac14$, but the integral can be defined by analytic continuation for all values $0<\eta<1$, and it is equal to the original Fourier integral (\ref{Cdef}) in the distribution sense. Provided $\eta$ is in this range, the result confirms the expected scaling $x_T\sim q_T^{-1}$ (since $e^{-\eta L_\perp}\propto(x_T^2\mu^2)^{-\eta}$) modulo a numerical ($\eta$-dependent) factor. Contributions from large values $x_T\gg q_T^{-1}$ are suppressed due to the rapid oscillations of the Bessel function, while contributions from small values $x_T\gg q_T^{-1}$ are phase-space suppressed. Additional powers of $L_\perp$ under the integral in (\ref{toy}), which enter when the ${\cal O}(a_s)$ corrections in (\ref{Cfinal}) are included, can be generated by taking derivatives with respect to $\eta$. It follows that the scale choice $\mu\sim q_T$ indeed eliminates large logarithms in the (Fourier-transformed) coefficient functions, as might have been expected from the beginning. 

The situation changes, however, at small transverse momentum. There are two possibilities that must be differentiated. Consider first the case where $\eta$ defined in (\ref{etadef}) remains smaller than 1 when $\mu\sim q_T$ approaches the non-perturbative domain. In this case the above discussion remains valid, but the perturbative calculation of the kernels $I_{q\leftarrow i}$ is no longer under theoretical control. Indeed, in this case one must resort to a formula for the cross section in terms of the transverse-position dependent PDFs $B_{i/N}(\xi,x_T^2,\mu)$ in (\ref{OPE}), which are then genuine, non-perturbative objects. But there is also a second possibility, that $\eta$ reaches 1 for values of $\mu$ that are still in the perturbative domain. We denote by $q_*$ the value of $\mu$ where this happens, i.e.\
\begin{equation}\label{qstar}
   q_* = M_Z\exp\left( - \frac{2\pi}{\Gamma_0^F\alpha_s(q_*)} \right) 
   \approx M_Z\exp\left( - \frac{2\pi}{\left(\Gamma_0^F+\beta_0\right)\alpha_s(M_Z)} \right) ,
\end{equation}
where in the last step we have used the one-loop approximation for the running coupling. Solving the first equation numerically, one finds $q_*\approx 1.88$\,GeV in the present case, which is indeed a reasonable short-distance scale. For $\eta\ge 1$, the integral in (\ref{toy}) is ultra-violet (UV) divergent for $x_T\to 0$, since the integrand then approaches $(b_0^2/\mu^2)^\eta\,x_T^{1-2\eta}$. It is then necessary to keep higher-order terms in the perturbative series for the exponent $g_F$ in (\ref{gFexp}). Indeed, since the quadratic term in $L_\perp^2$ has a negative coefficient, it provides a gaussian weight to the integral which cuts off the divergence at $x_T\to 0$. But at the same time, the quadratic term also provides a regulator for the infrared (IR) region of very large $x_T$, which is in addition to the oscillating behavior of the Bessel function. For small $q_T<q_*$, this gaussian fall-off is the dominating factor, which prevents that $x_T$ can become arbitrarily large. Remarkably, this implies that $\langle x_T\rangle$ decouples from $q_T^{-1}$ and stays in the short-distance domain even in the extreme case where $q_T$ is taken to 0. Using (\ref{gFexp}) and changing variables from $x_T$ to $\ell=L_\perp$, we obtain
\begin{eqnarray}\label{Cq0}
   \bar C_{q\bar q\leftarrow ij} \big|_{q_T\to 0}
   &=& \frac{b_0^2}{4\mu^2} \int_{-\infty}^\infty\!d\ell\,\exp\left[ (1-\eta)\,\ell
    - a_s \left[ \left( \Gamma_0^F + \eta\beta_0 \right) \frac{\ell^2}{2}   
    + \left( 2\gamma_0^q + \eta K \right) \ell + \eta d_2 \right]
    + {\cal O}(a_s^2) \right] \nonumber\\
   &&\times \bar I_{q\leftarrow i}(z_1,\ell,a_s)\,
    \bar I_{\bar q\leftarrow j}(z_2,\ell,a_s) \,.
\end{eqnarray}
The integrand features a gaussian peak at
\begin{equation}
   \ell_{\rm peak} 
   = \frac{1-\eta-a_s\left( 2\gamma_0^q + \eta K \right)}{a_s\left( \Gamma_0^F + \eta\beta_0 \right)} 
\end{equation}
with a width proportional to $1/\sqrt{a_s}$. The condition that at the peak the logarithm $\ell=L_\perp$ should be an ${\cal O}(1)$ quantity implies that $1-\eta={\cal O}(a_s)$, implying that the factorization scale must be chosen in the vicinity of the scale $q_*$ in (\ref{qstar}). In other words, once the scale $\mu$ reaches $q_*$, the value of $\mu$ that keeps the logarithms $L_\perp$ small decouples from $q_T$ and stays near $q_*$ even when $q_T\to 0$: 
\begin{equation}
   \mu\sim\langle x_T^{-1}\rangle\sim \max( q_T, q_* ) \,.
\end{equation}
In our numerical work, we will use $\mu=q_T+q_*$ as the default choice. We recall that the emergence of the scale $q_*$ in (\ref{Cdef}), below which the scaling of $x_T^{-1}$ decouples from $q_T$, is a consequence of the collinear anomaly, which is responsible for the anomalous dependence on $M_Z$ in (\ref{Cdef}). Provided the mass of the Drell-Yan boson is large enough that $q_*\gg\Lambda_{\rm QCD}$, the transverse-momentum distribution is protected from long-distance physics even for arbitrarily small $q_T$. The resummed perturbative series for the cross section generates the scale $q_*$ dynamically, and even though this is a short-distance scale, it is related to the boson mass $M_Z$ in a genuinely non-perturbative way.

The above discussion shows that we must distinguish two regions of transverse momenta. For $q_T\gg q_*$ the Bessel function regularizes the UV region, and the scale choice $\mu\sim q_T$ prevents that the logarithms $L_\perp$ give rise to large perturbative corrections. It is then consistent to count these logarithms as $L_\perp\sim 1$ and construct the perturbative series as a series in powers of $a_s$. For $q_T\ll q_*$ the situation is different. Even though the scale choice $\mu\sim q_*$ ensures that $L_\perp={\cal O}(1)$ at the peak of the integrand, the gaussian weight factor allows for significant contributions to the integral over a range of $L_\perp$ values with width proportional to $1/\sqrt{a_s}$, see (\ref{Cq0}). It follows that for very small $q_T$ the resummation procedure must be reorganized, using the modified power counting $L_\perp\sim 1/\sqrt{a_s}$. This implies that single-logarithmic terms $\left(a_s L_\perp\right)^n\sim a_s^{n/2}$ are always suppressed, whereas double-logarithmic terms $\left(a_s L_\perp^2\right)^n\sim 1$ are unsuppressed and must be resummed to all orders. To keep track of this fact, we introduce an auxiliary expansion parameter $\epsilon$ (which at the end is set to 1) and assign the power counting $a_s\sim\epsilon$ and $L_\perp\sim\epsilon^{-1/2}$. In Appendix~\ref{app:b}, we use the recursive solutions to the RG equations (\ref{Fevol}) and (\ref{hevol}) to determine all terms in $F_{q\bar q}$ and $h_F$ that contribute up to ${\cal O}(\epsilon)$ to the exponent $g_F$ defined in (\ref{gdef}). This involves some four-loop contributions to $F_{q\bar q}$ and some three-loop contributions to $h_F$, which however can all be expressed in terms of one- and two-loop coefficients of the anomalous dimensions and $\beta$-function. The resulting expression is
\begin{eqnarray}\label{gFmod}
   g_F(\eta,L_\perp,a_s) 
   &=& - \big[ \, \eta L_\perp \, \big]_{\epsilon^{-1/2}}
    - \left[ a_s \left( \Gamma_0^F + \eta\beta_0 \right) \frac{L_\perp^2}{2}
    \right]_{\epsilon^0} \nonumber\\
   &&\mbox{}- \left[ a_s \left( 2\gamma_0^q + \eta K \right) L_\perp
    + a_s^2 \left( \Gamma_0^F+ \eta\beta_0 \right) \beta_0\,\frac{L_\perp^3}{3} 
    \right]_{\epsilon^{1/2}} \\
   &&\mbox{}- \left[ a_s\,\eta d_2
    + a_s^2 \Big( K\Gamma_0^F + 2\gamma_0^q\beta_0 + \eta\,\big( \beta_1 + 2K\beta_0 \big) 
    \Big) \frac{L_\perp^2}{2}
    + a_s^3 \left( \Gamma_0^F + \eta\beta_0 \right) \beta_0^2\,\frac{L_\perp^4}{4} 
    \right]_{\epsilon} \nonumber\\
   &&\mbox{}- {\cal O}(\epsilon^{3/2}) \,, \nonumber
\end{eqnarray}
which is more complicated than the naive perturbative expansion in (\ref{gFexp}). The auxiliary parameter $\epsilon$ counts the order in $a_s$ resulting (for $q_T\ll q_*$) after the $x_T$ integral in (\ref{Cfinal}) has been performed. The two terms given in the first line are unsuppressed and must be kept in the exponent of the integral in (\ref{Cfinal}), whereas the remaining terms can be expanded in powers of $\epsilon^{1/2}$. The resulting integrals over the Bessel function in (\ref{Cfinal}) can readily be evaluated numerically. An efficient way of doing this is to use that $J_0(x_T q_T)=\frac{2}{\pi}\,\mbox{Im}\,K_0(-i x_T q_T)$ and to perform a contour rotation from $x_T\to i x_T$.

It is interesting to ask whether the ``protective behavior'' in both the UV and IR regions provided by the gaussian terms in the exponent could be upset at yet higher orders in the perturbative expansion of the exponent. It is not difficult to show that the highest-order logarithmic terms in (\ref{gFmod}) are given to all orders by $-a_s^{n+1}\,(\Gamma_0^F+\eta\beta_0)\,\beta_0^n\,L_\perp^{n+2}/(n+2)$ with $n\ge 0$. Provided that the series is truncated at an integer order in $\epsilon$, in which case the largest $n$ value is even, these terms are negative, and they guarantee that the integral over $x_T$ in (\ref{Cfinal}) converges both in the UV and IR regions. When the expansion is performed in the exponent, the series of these highest logarithms does not exhibit a factorial divergence. On the contrary, the sum is convergent as long as $L_\perp<1/(\beta_0 a_s)$, which is always the case with our power counting $L_\perp\sim 1/\sqrt{a_s}$. Hence, the modified power counting displayed in (\ref{gFmod}) indeed provides a consistent, well-behaved expansion scheme for the integral over the Bessel function.

We finally give the expressions for the collinear kernel function arising with our modified power counting. We find
\begin{equation}\label{Imod}
\begin{aligned}
   \bar I_{q\leftarrow i}(z,L_\perp,a_s) 
   &= \delta(1-z)\,\delta_{qi} - \left[ a_s\,{\cal P}_{q\leftarrow i}^{(1)}(z)\,
    \frac{L_\perp}{2} \right]_{\epsilon^{1/2}} \\
   &\quad\mbox{}+ \left[ a_s\,{\cal R}_{q\leftarrow i}(z) 
    + a_s^2 \left( {\cal D}_{q\leftarrow i}(z) - 2\beta_0\,{\cal P}_{q\leftarrow i}^{(1)}(z)
    \right) \frac{L_\perp^2}{8} \right]_\epsilon + {\cal O}(\epsilon^{3/2}) \,,
\end{aligned}
\end{equation}
which may be compared with (\ref{barIexp}). Here 
\begin{equation}\label{Ddef}
   {\cal D}_{q\leftarrow i}(z) 
   = \sum_{j=q,g} \int_z^1\!\frac{du}{u}\,{\cal P}_{q\leftarrow j}^{(1)}(u)\,
    {\cal P}_{j\leftarrow i}^{(1)}(z/u)
\end{equation}
involve the convolutions of two DGLAP splitting functions. The resulting expressions are given in  Appendix~\ref{app:b}.

Formulas (\ref{gFmod}) and (\ref{Imod}) are our main results. With the help of these expressions, large logarithms can be resummed at NNLL order all the way down to zero transverse momentum (always assuming that $q_*$ is in the perturbative domain, as is indeed the case for $Z$-boson production). For larger $q_T$ values the additional terms contained in (\ref{gFmod}) and (\ref{Imod}) compared with (\ref{gFexp}) and (\ref{barIexp}) reduce to higher-order terms proportional to $a_s^2$ and $a_s^3$, which can be neglected to the order we are working. Hence, our formula provides a smooth interpolation between the regions of small and very small $q_T$.

Interestingly, the additional terms needed at very low $q_T$ are also important at $q_T>q_*$, as we will now show. We have argued above that the quadratic terms in $L_\perp$ need to be kept in the exponent as $\eta$ approaches 1, in order to regularize the integral over $x_T$ in the UV region. As long as $\eta$ is less than 1 (i.e.\ for $q>q_*$), it would seem justified to expand out the ${\cal O}(a_s)$ correction to the exponent $g_F$ in (\ref{gFexp}) in a power series in $a_s$. Surprisingly, it turns out that this would be a bad idea. Closer inspection of the integral in (\ref{toy}) shows that even for $(1-\eta)$ small but not zero, taking $n$ derivatives with respect to $\eta$ (corresponding to an extra factor $(-L_\perp)^n$ under the integral) generates a contribution exhibiting factorial growth in $n$. For the special case $\mu=q_T$ the series has been analyzed in \cite{Becher:2010tm}, where it was shown that (setting $a\equiv 2a_s(\Gamma_0^F+\eta\beta_0)$ for brevity)
\begin{equation}\label{Besselint}
\begin{aligned}
   &\quad\frac12 \int_0^\infty\!dx_T\,x_T\,J_0(x_T q_T)\,
    e^{-\eta L_\perp-\frac{a}{4} L_\perp^2} \Big|_{\mu=q_T,\,\,\eta\lesssim 1}^{\rm expanded} \\
   &= \frac{e^{-2\gamma_E}}{q_T^2}\,\sum_{n=0}^\infty \left\{
    \frac{(2n)!}{n!} \left( - \frac{a}{4} \right)^n
    \left[ \frac{1}{(1-\eta)^{2n+1}} - e^{-2\gamma_E} \right] 
    + k_n\,a^n + {\cal O}(1-\eta) \right\} ,
\end{aligned}
\end{equation}
where the coefficients $k_n$ do not exhibit the strong factorial growth of the terms in the first sum. (Numerically, we find $k_0\approx 0.315$, $k_1\approx 0.243$, $k_2\approx-0.075$, $k_3\approx-0.051$, $k_4\approx 0.045$, \dots.) While the series in (\ref{Besselint}) is badly divergent, the fact that it has alternating sign implies that it can be Borel summed, yielding
\begin{equation}\label{I1Borel}
\begin{aligned}
   &\quad\frac12 \int_0^\infty\!dx_T\,x_T\,J_0(x_T q_T)\,
    e^{-\eta L_\perp-\frac{a}{4} L_\perp^2} \Big|_{\mu=q_T,\,\,\eta\lesssim 1}^{\rm Borel~sum} \\
   &= \frac{e^{-2\gamma_E}}{q_T^2}\,\sqrt{\frac{\pi}{a}} \left\{ e^{\frac{(1-\eta)^2}{a}} 
    \left[ 1 - \mbox{Erf}\left( \frac{1-\eta}{\sqrt{a}} \right) \right]
    - e^{-2\gamma_E+\frac1a} \left[ 1 - \mbox{Erf}\left( \frac{1}{\sqrt{a}} \right) 
    \right] \right\} \\
   &\quad\mbox{}+ \frac{e^{-2\gamma_E}}{q_T^2}\,\sum_{n=0}^\infty\,k_n\,a^n + {\cal O}(1-\eta) \,,
\end{aligned}
\end{equation}
where $\mbox{Erf}(x)$ is the error function. This expression provides an explicit (albeit approximate) result for the complicated integral on the left-hand side. Note that the spurious singularity at $\eta=1$ in the sum (\ref{Besselint}) has disappeared after Borel summation, but that the final answer depends in a highly non-perturbative way on the coupling constant $a_s\sim a$. Even though a perturbative expansion in powers of $a\sim a_s$ can be constructed, its radius of convergence is zero. The lesson from this exercise is that even for values of $\eta$ large but not very close to 1 it is obligatory to keep the ${\cal O}(a_s L_\perp^2)$ terms in the exponent of the $x_T$ integral, as suggested by our modified power counting in (\ref{gFmod}). Expanding these terms out is only justified for rather large $q_T$, for which $\eta={\cal O}(\alpha_s)$ is a small quantity.

How can this strange behavior be understood? The integral over the $q_T$ spectrum from 0 up to some value $q_T^{\rm max}\gg\Lambda_{\rm QCD}$ should be calculable in fixed-order perturbation theory, as is the integral over all possible values. Why do problems arise for smaller values of $q_T$, which naively are still in the perturbative domain? As we will show in the following section, the intercept of the spectrum $d\sigma/dq_T^2$ (or $d^2\sigma/dq_T^2 dy$) can be calculated using short-distance methods, but it exhibits an essential singularity at $\alpha_s=0$ and hence does not admit any perturbative expansion. Since the spectrum is at least approximately given by a smoothly falling function, the area under it (the total cross section) is roughly proportional to the height (intercept) times a typical width. The requirement that the area be perturbatively calculable requires that not only the height, but also the characteristic width of the spectrum are governed by non-perturbative short-distance physics.

\begin{figure}
\begin{center}
\begin{tabular}{c}
\includegraphics[width=0.95\textwidth]{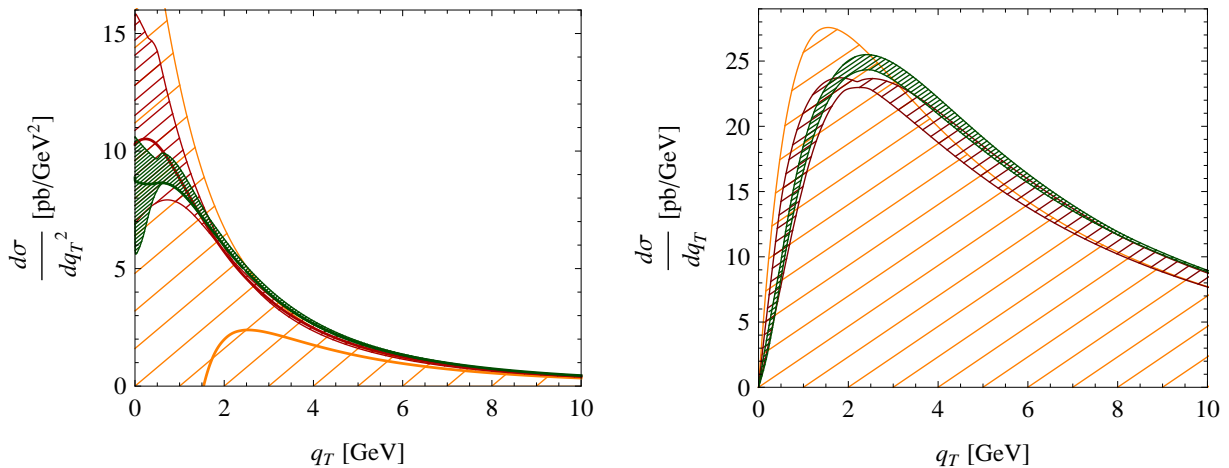}
\end{tabular}
\vspace{-1.0cm}
\end{center}
\caption{\label{fig:3regions}
Comparison of different expansion schemes. The bands result from varying $\mu$ by a factor two around the default value $\mu=q_T+q_*$. The thick lines in the left plot show the default predictions. The wide orange bands are obtained using the naive resummation formula, which suffers from a factorial divergence. The red bands result if the divergent terms are resummed (conventional resummation). The dark green, densely hatched bands arise if one further resums the terms which are enhanced at very small $q_T$ (improved expansion).}
\end{figure}

In order to make these comments more precise and see the relevance of the different contributions discussed in this section, we show in Figure~\ref{fig:3regions} the differential cross sections $d\sigma/dq_T^2$ (left) and $d\sigma/dq_T$ (right) for $Z$-boson production (with subsequent decay to a lepton pair $Z\to \ell^+\ell^-$) at the Tevatron, using three different approximation schemes. 
The green bands show the cross sections at NNLL order in our improved expansion scheme based on the
modified power counting given by the $\epsilon$ expansion discussed above (``improved resummation''). The red bands show results obtained with a conventional power counting in $a_s$, keeping however the quadratic terms of order $a_s L_\perp^2$ in the exponent (``conventional resummation''). Finally, the orange bands correspond to the resummation scheme where all $a_s$ terms in the exponent are expanded out (``naive resummation''). Considering first the plot on the left, we see that as expected the improved resummation becomes important for $q_T\lesssim q_*\approx 1.88$\,GeV, where it gives results that are lower and have a smaller scale uncertainty compared with the conventional resummation scheme. In contrast, the naive expansion, which is affected by the factorial divergence, leads to very large scale dependence, such that the cross section becomes negative if the scale is lowered by a factor of two from its default value, and even the default prediction shown by the solid thick orange curve  becomes negative for $q_T<1.5\,{\rm GeV}$. In the region of very small transverse momentum, the conventional resummation scheme is only valid to NLL accuracy, since it misses the additional higher-order terms which are enhanced in this region. As a consequence, the scale uncertainty near $q_T=0$ is much larger in this case than for the improved resummation scheme. Interestingly, the effects of including the additional terms is also non-negligible at higher values of $q_T$. While the results obtained in the conventional and improved resummation schemes are compatible within scale uncertainties, the improved result has a significantly smaller uncertainty, and the peak of the distribution is shifted slightly to the right. The reason that the effects of the additional resummation are still visible away from the end-point is the strong fall-off of the spectrum towards larger $q_T$.

In addition to the logarithmic terms which are resummed by our result, the cross section also contains regular terms, which can be obtained from a fixed-order computation of the spectrum. In order to capture both corrections, we combine our result with the fixed-order result for the $q_T$ spectrum. To avoid double counting of the logarithmic terms, we need to subtract the fixed-order expansion of our resummed result from the full fixed-order result. To obtain a result which is valid both to NNLL and at NLO in fixed-order perturbation theory, we compute
\begin{equation}
   \frac{d\sigma^{\rm NNLL}}{dq_T} \bigg|_{\rm matched}
   = \frac{d\sigma^{\rm NNLL}}{dq_T} + \left( \frac{d\sigma^{\rm NLO}}{dq_T}
    - \frac{d\sigma^{\rm NNLL}}{dq_T} \right) \bigg|_{\text{expanded to NLO}} \,.
\end{equation}
The $q_T$ spectrum is known to ${\cal O}(\alpha_s^2)$ \cite{Ellis:1981hk,Arnold:1988dp,Gonsalves:1989ar}, but it turns out that the matching corrections are tiny in the peak region, and ${\cal O}(\alpha_s)$ matching is sufficient for our purposes. To obtain the fixed-order expansion of the resummed result, we evaluate the hard matching coefficient $\left|C_V(-M_Z^2,\mu)\right|^2$ in fixed-order perturbation theory, setting $\mu_h=\mu$ in relations (\ref{CVsol}) and (\ref{CV2fopt}) of Appendix~\ref{app:hard}, and expand the exponent $g_F$ in powers of $\alpha_s$. After this expansion, the integrals over the Bessel function in (\ref{Cfinal}) can be computed analytically. A convenient way to do this is to first keep the term $\eta L_\perp$ term in the exponent, and then use (\ref{toy}) to obtain the Fourier integral. The Fourier transform of the higher-order logarithmic terms can be obtained by taking derivatives with respect to $\eta$ of relation (\ref{toy}). The resulting expression was given explicitly in equation (60) of \cite{Becher:2010tm}. The result for the ${\cal O}(\alpha_s^2)$ expansion of our result is given in Appendix~\ref{nloexpansion}.

\section{Intercept and behavior near $\bm{q_T=0}$}
\label{sec:intercept}

It is an interesting exercise to use the general results of the previous section to derive an analytic expression for the intercept of the differential cross section $d^2\sigma/dq_T^2\,dy$ at $q_T=0$. On one side, this will give us a better understanding of the structure of resummed perturbation theory for very small $q_T$. Indeed, we will obtain an expression for the intercept that exhibits an essential singularity at $a_s=0$, meaning that not even a divergent perturbative expansion can be constructed. On the other hand, precise control over the value of the intercept might also be of phenomenological relevance in view of future precision measurements at the LHC, as it might help to constrain and extrapolate the data into the difficult region of very small transverse momentum. It is a curious fact that more than 30 years after the discovery that the intercept is calculable provided that the Drell-Yan mass is very large \cite{Parisi:1979se}, there is still no explicit expression in the literature for the ${\cal O}(1)$ normalization factor of the result. The saddle-point evaluations performed in \cite{Parisi:1979se,Collins:1984kg,Ellis:1997ii} provide the leading asymptotic behavior for $M_Z\to\infty$ only, without fixing the normalization of the cross section. Our effective field-theory approach allows us to not only derive a closed expression for the intercept at leading order in resummed perturbation theory, but we are also in a position to show that corrections to this result can be calculated in a power series in $\alpha_s$. The first-order correction will be given explicitly below.

For $q_T=0$ the relevant integral over transverse displacement takes the form (\ref{Cq0}), where in the exponent we must substitute the expression (\ref{gFmod}) for $g_F$. Using that $1-\eta={\cal O}(a_s)={\cal O}(\epsilon)$ in this case, we can rewrite this expression as
\begin{eqnarray}\label{gFmodexp}
   g_F(\eta\approx 1,L_\perp,a_s) 
   &=& - \big[ \, L_\perp \, \big]_{\epsilon^{-1/2}}
    - \left[ a_s \left( \Gamma_0^F + \beta_0 \right) \frac{L_\perp^2}{2}
    \right]_{\epsilon^0} \nonumber\\
   &&\mbox{}- \left[ 
    a_s \bigg( 2\gamma_0^q + K + (\Gamma_0^F+\beta_0) \ln\frac{q_*^2}{\mu^2} \bigg) L_\perp
    + a_s^2 \left( \Gamma_0^F + \beta_0 \right) \beta_0\,\frac{L_\perp^3}{3} 
    \right]_{\epsilon^{1/2}} \nonumber\\
   &&\mbox{}- \left[ a_s\,d_2
    + a_s^2 \bigg( K\Gamma_0^F + 2\gamma_0^q\beta_0 + \beta_1 + 2K\beta_0  
    + (\Gamma_0^F+\beta_0)\,\beta_0 \ln\frac{q_*^2}{\mu^2} \bigg) \frac{L_\perp^2}{2} \right. 
    \nonumber\\
   &&\hspace{6mm}\left. \mbox{}+ a_s^3 \left( \Gamma_0^F + \beta_0 \right) \beta_0^2\, 
    \frac{L_\perp^4}{4} \right]_{\epsilon} - {\cal O}(\epsilon^{3/2}) \,.
\end{eqnarray}
Here $\ln(q_*^2/\mu^2)$ counts as ${\cal O}(1)$ and allows for scale choices in the vicinity of the default value $\mu=q_*$. The formally super-leading ${\cal O}(L_\perp)$ term is absorbed by the change of variables from $x_T$ to $\ell=L_\perp$, while the leading ${\cal O}(L_\perp^2)$ term needs to be kept in the exponent and provides the gaussian weight factor. Expanding out the higher-order terms in (\ref{Imod}) and (\ref{gFmodexp}) then gives rise to gaussian integrals, which can readily be evaluated. In this way, we obtain at next-to-leading order in $a_s$
\begin{eqnarray}\label{Cq0fin}
   \bar C_{q\bar q\leftarrow ij}(z_1,z_2,0,M_Z^2,\mu)
   &=& \frac{e^{-2\gamma_E}}{\mu^2}\,
    \sqrt{\frac{2\pi}{a_s \left( \Gamma_0^F + \beta_0 \right)}}\,\,\Bigg\{
    \Big[ 1 + c_1(\mu)\,a_s \Big]\,\delta(1-z_1)\,\delta(1-z_2)\,
    \delta_{qi}\,\delta_{\bar qj} \nonumber\\
   &&\mbox{}+ a_s \bigg[
    \left( c_2(\mu)\,{\cal P}_{q\leftarrow i}^{(1)}(z_1) + {\cal R}_{q\leftarrow i}(z_1)
    + \frac{{\cal D}_{q\leftarrow i}(z_1)}{8\left(\Gamma_0^F+\beta_0\right)} \right)
    \delta(1-z_2)\,\delta_{\bar qj} \nonumber\\
   &&\hspace{11mm}\mbox{}
    + \frac{{\cal P}_{q\leftarrow i}^{(1)}(z_1)\,{\cal P}_{\bar q\leftarrow j}^{(1)}(z_2)}%
           {8\left(\Gamma_0^F+\beta_0\right)}
    + (q,i,z_1\leftrightarrow\bar q,j,z_2) \bigg] \Bigg\} \,, 
\end{eqnarray}
where for $n_f=4$ 
\begin{equation}
\begin{aligned}
   c_1(\mu) 
   &= \frac{\left(K+2\gamma_0^q\right)^2}{2\left(\Gamma_0^F+\beta_0\right)}
    - \frac{K\Gamma_0^F+\beta_1-2\gamma_0^q\beta_0}{2\left(\Gamma_0^F+\beta_0\right)}
    + \frac{\beta_0^2}{12\left(\Gamma_0^F+\beta_0\right)} - d_2 \\
   &\quad\mbox{}+ \left( K + 2\gamma_0^q + \frac{\beta_0}{2} \right) \ln\frac{q_*^2}{\mu^2}
    + \frac{\Gamma_0^F+\beta_0}{2}\,\ln^2\frac{q_*^2}{\mu^2} \\
   &= 1.48854 + 4.18595 \ln\frac{q_*^2}{\mu^2} + 6.83333 \ln^2\frac{q_*^2}{\mu^2} \,, \\
   c_2(\mu) &= \frac{K+2\gamma_0^q+\frac12\,\beta_0}{2\left(\Gamma_0^F+\beta_0\right)} 
    + \frac12 \ln\frac{q_*^2}{\mu^2} 
    = 0.153145 + 0.5 \ln\frac{q_*^2}{\mu^2} \,.
\end{aligned}
\end{equation}
Note the peculiar, power-like dependence of the result (\ref{Cq0fin}) on the factorization scale. Using that for $\mu$ in the vicinity of $q_*$ we can rewrite
\begin{equation}
   \Gamma_{\rm cusp}^F(a_s)\,\ln\frac{M_Z^2}{\mu^2}
   = 1 + K a_s + \left( \Gamma_0^F+\beta_0 \right) a_s\,\ln\frac{q_*^2}{\mu} + {\cal O}(a_s^2)
\end{equation}
in (\ref{Cevol}), it is straightforward to check that the product $|C_V(-M_Z^2,\mu)|^2\,\bar C_{q\bar q\leftarrow ij}(z_1,z_2,0,M_Z^2,\mu)$ is indeed RG invariant. Numerically, the NLO correction proportional to $a_s$ in (\ref{Cq0fin}) is of rather modest size. 

With the default scale choice $\mu=q_*$, the prefactor in (\ref{Cq0fin}) can be written as
\begin{equation}
   \frac{e^{-2\gamma_E}}{\mu^2}\,\sqrt{\frac{2\pi}{a_s \left( \Gamma_0^F + \beta_0 \right)}}\,
    \Bigg|_{\mu=q_*}
   = \frac{2\pi e^{-2\gamma_E}}{M_Z^2}\,\exp\left[ \frac{\pi}{C_F\alpha_s(q_*)}\right]
    \sqrt{\frac{2}{\left( \Gamma_0^F + \beta_0 \right) \alpha_s(q_*)}} \,,
\end{equation}
which features an essential singularity at $\alpha_s=0$. When combined with the NLO expression for the hard function $\left|C_V(-M_Z^2,\mu)\right|^2$ given in Appendix~\ref{app:hard}, the result (\ref{Cq0fin}) provides an explicit expression for the intercept of the Drell-Yan spectrum $d^2\sigma/dq_T^2\,dy$ at $q_T=0$, to NLO in RG-improved perturbation theory. While the leading term is of a genuinely non-perturbative nature, higher-order corrections can be calculated in a systematic way in powers of $\alpha_s(M_Z)$ (for $|C_V|^2$) and $\alpha_s(q_*)$ (for $\bar C_{q\bar q\leftarrow ij}$). Even though the the calculability of the intercept (for sufficiently large Drell-Yan mass) is known since the paper \cite{Parisi:1979se}, to the best of our knowledge this is the first time that explicit expressions for the normalization and the first-order perturbative correction have been derived. 

Since resummed perturbation theory allows one to predict the intercept of the transverse-momentum spectrum at $q_T=0$, it is natural to ask whether similar methods can be employed to construct a series expansion of the differential cross section $d^2\sigma/dq_T^2 dy$ in powers of $q_T^2$, which is valid for $q_T<q_*$. To this end, one would like to compute derivatives of the distribution (\ref{Cfinal}) with respect to $q_T^2$ evaluated at $q_T=0$. Any attempt to do so leads to an encounter with a violently divergent series. Taking $n$ derivatives of the Bessel function $J_0(x_T q_T)$ with respect to $q_T^2$ and setting $q_T=0$ generates a factor $x_T^{2n}\sim e^{n\ell}$ in the integrand in (\ref{Cq0}), and evaluating the resulting gaussian integral for $\eta\approx 1$ yields an extra factor of $e^{n^2/[2a_s (\Gamma_0^F+\beta_0)]}$. The leading term in (\ref{Cq0fin}) is then multiplied by the series
\begin{equation}\label{crash}
\begin{aligned}
   & \sum_{n=0}^\infty\,\frac{(-1)^n}{e^{2n\gamma_E}\left(n!\right)^2} 
    \left( \frac{q_T^2}{q_*^2} \right)^n
    \exp\left[ \frac{2\pi n^2}{\left(\Gamma_0^F+\beta_0\right) \alpha_s(q_*)} \right] \\
   &\approx 1 - 1.435\,\frac{q_T^2}{q_*^2} + 10.66 \left( \frac{q_T^2}{q_*^2} \right)^2
    - 729.7 \left( \frac{q_T^2}{q_*^2} \right)^3 
    + 5.82\cdot 10^5 \left( \frac{q_T^2}{q_*^2} \right)^4 \mp \dots \,.
\end{aligned}
\end{equation}
The resulting divergence is much worse than a factorial growth in $n$ and renders any attempt to derive an analytic expression for the shape of the $q_T$ spectrum near the origin hopeless. Having already seen that the integral over the Bessel function in the first line of (\ref{Besselint}) does not want to be expanded in powers of $a_s$, we now observe that, even more so, it does not want to be expanded in powers of $q_T$. The shape of the distribution for small transverse momenta is genuinely non-analytic.

The discussion just presented raises an interesting issue. In (\ref{OPE}), we have presented the first term in an OPE of the transverse-position dependent PDFs $B_{i/N}$ in terms of ordinary PDFs defined in terms of the nucleon matrix element of a bilocal quark operator at zero transverse separation. In SCET, it would be straightforward to extend this expansion to higher orders in $\Lambda_{\rm QCD}^2\,x_T^2$. For very small $q_T$ we would expect the dynamically generated scale $q_*$ to determine the size of these power corrections. However, at any {\em finite\/} order such an OPE will run into the same problem as mentioned above, generating ``power corrections'' whose coefficients grow like $e^{\# n^2/\alpha_s}$. In order to avoid this problem the OPE must be resummed, which of course is not feasible in practice. On the other hand, reasonable models for such a resummed OPE can be obtained by noting that the transverse-position dependent PDFs must vanish rapidly when the two quark fields are separated by a transverse distance $x_T$ larger than the proton size, since then the quark bilinear has very little overlap with the external nucleon state. It is thus reasonable to use an ansatz of the form
\begin{equation}\label{Binonpert}
   B_{q/N}(\xi,x_T^2,\mu)
   = f_{\rm hadr}(x_T\Lambda_{\rm NP})\,B_{q/N}^{\rm pert}(\xi,x_T^2,\mu)\, \,,
\end{equation}
where the perturbative functions $B_{i/N}^{\rm pert}$ carry all the scale dependence and are given by (\ref{OPE}), whereas the hadronic form factor $f_{\rm hadr}(r)$ with $f_{\rm hadr}(0)=1$ describes the fall-off at large transverse distances and is parameterized in terms of a hadronic scale $\Lambda_{\rm NP}$. For simplicity, we assume that this form factor is independent of $\xi$. The above ansatz inserts a factor $[f_{\rm hadr}(x_T\Lambda_{\rm NP})]^2$ under the integral over $x_T$ in (\ref{Cfinal}), which suppresses the region of very large $x_T$ values. We will adopt the models
\begin{equation}\label{fmodels}
   f_{\rm hadr}^{\rm gauss}(x_T\Lambda_{\rm NP}) = \exp\left( -\Lambda_{\rm NP}^2\,x_T^2 \right) ,
    \qquad
   f_{\rm hadr}^{\rm pole}(x_T\Lambda_{\rm NP}) 
   = \frac{1}{1+\Lambda_{\rm NP}^2\,x_T^2}
\end{equation}
for the form factor, which agree in their first-order terms but have quite different behavior for large separation. Fortunately, we will find that while the results are rather sensitive to the value of the hadronic scale $\Lambda_{\rm NP}$, the precise shape of the form factor appears to be of minor importance. We will show in Section~\ref{sec:numerics} that hadronic corrections to the value of the intercept of the $d\sigma/dq_T^2$ distribution indeed scale (approximately) as a power law, $\sim(\Lambda_{\rm QCD}/q_*)^\delta$, where due to the resummation of the OPE the exponent $\delta$ is not given by an even integer. Let us note for completeness that the hadronic form factor $f_{\rm hadr}$ can in general also depend on the quark flavor and the momentum fraction $\xi$. Models which include a factor $\xi^{\Lambda_2^2 x_T^2 }$ in $f_{\rm hadr}$, with a second non-perturbative parameter $\Lambda_2$, were studied in \cite{Ladinsky:1993zn,Konychev:2005iy}.

\section{Systematic studies}
\label{sec:numerics}

Having discussed the structure of the theoretical prediction for the resummed Drell-Yan cross section in detail, we now proceed to perform some systematic studies related to scale variations, various implementations of the expansion, different ways to set the scale $\mu$, and the importance of power corrections. For the purposes of the discussion in this section, we will consider the resummed cross section $d\sigma/dq_T$ without matching to fixed-order perturbation theory. We show results at LO and NLO in RG-improved perturbation theory, which correspond to NLL and NNLL accuracy. Since we will later match to fixed-order calculations, we will refer in the following to the resummed results by their logarithmic accuracy rather than their order in RG-improved perturbation theory so as to avoid confusion. For concreteness, we consider the case of $Z$-boson production at the Tevatron with the $Z$-boson decaying leptonically, $p\bar{p}\to X+ Z \to X+\ell^+\ell^-$. The corresponding cross section is obtained by multiplying the $Z$-production cross section with the leptonic branching ratio ${\rm Br}(Z\to\ell^+\ell^-)=0.03366$. Throughout, we use $\sin^2\theta_W=0.2312$ for the weak mixing angle and $\alpha(M_Z)=1/128.89$ for the fine-structure constant. We use MSTW2008NNLO \cite{Martin:2009iq} as our default PDF set, which has an associated value of $\alpha_s(M_Z)=0.11707$. The strong coupling is evolved with three-loop accuracy and has flavor thresholds at $\mu_b=4.75$\,GeV and $\mu_c=1.4$\,GeV for the $b$ and $c$ quarks.

\begin{figure}[t!]
\begin{center}
\includegraphics[width=0.95\textwidth]{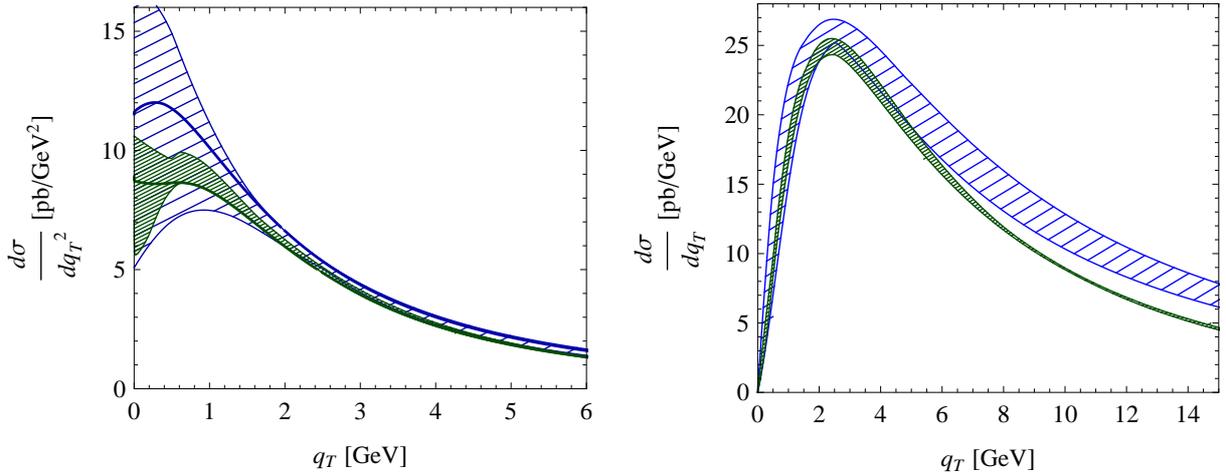} 
\end{center}
\vspace{-1.0cm}
\caption{\label{fig:NLLvsNNLL}
Comparison of NLL (blue bands) and NNLL (green bands) predictions for the cross section in the improved expansion scheme. The factorization scale $\mu$ is varied by a factor two about its default value $\mu=q_T+q_*$, while the hard matching scale is fixed at $\mu_h^2=-M_Z^2$. The thick lines in the left plot are obtained for the default scale choice.}
\end{figure}

\begin{table}[t!]
\begin{center}
\vspace{0.3cm}
\begin{tabular}{lcc}
 & $\mu_h^2=m_Z^2$ &  $\mu_h^2=-m_Z^2$  \\ \hline
NLL & $1.000^{+0.160}_{-0.060}$ & $1.334^{+0.201}_{-0.074}$ \\ 
NNLL & $1.087^{+0.010}_{-0.001}$ & $1.131^{+0.001}_{-0.014}$ \\ 
N$^3$LL & $1.119^{+0.006}_{-0.001}$ & $1.130^{+0.001}_{-0.001}$
\end{tabular}
\end{center}
\caption{\label{tabhard} 
The hard function $|C_V(-M_Z^2,\mu)|^2$ at $\mu=M_Z$ for space-like and time-like choices of $\mu_h^2$. The uncertainties are obtained by varying $\mu_h$ by a factor two about the default value.}
\end{table}

In Figure~\ref{fig:NLLvsNNLL}, we show the scale dependence of the cross section obtained in the improved resummation scheme developed in Section~\ref{sec:expansion}, varying the factorization scale $\mu$ by a factor two about the default choice $\mu=q_*+q_T$, where $q_*\approx 1.88$\,GeV has been defined in (\ref{qstar}) and emerges dynamically in the region of very small transverse momentum. We observe a significant reduction of the scale uncertainty when going from NLL to NNLL order. In addition to the scale $\mu$, the cross section also depends on the hard matching scale $\mu_h$, which arises when the hard function $|C_V(-M_Z^2,\mu)|^2$ is evolved from a high scale $\mu_h\sim M_Z$ to the scale $\mu$. The solution of the corresponding RG equation is well known and is reproduced in Appendix~\ref{app:hard}. If $\mu_c<\mu< \mu_b$, the hard function has to be evolved across a flavor threshold. In this case, we first evolve from $\mu_h$ down to the threshold $\mu_b$, switch to four flavors, and then evolve from $\mu_b$ to $\mu$. While the all-order solution is independent of the matching scale $\mu_h$, a residual scale dependence remains at finite orders. To estimate the associated uncertainty, we should vary the matching scale $\mu_h$ in the hard function. However, since the hard function is an overall factor multiplying the cross section, we can discuss the $\mu_h$ dependence independently from the rest. To separate off the $q_T$ dependence arising from the choice of $\mu$, we set $\mu=M_Z$ for the discussion of the $\mu_h$ dependence. The perturbative expansion of the hard function and its $\mu_h$ dependence are displayed in Table~\ref{tabhard}, which shows two different choices for the default matching scale: the space-like choice $\mu_h^2=M_Z^2$ and the time-like choice $\mu_h^2=-M_Z^2$. Picking $\mu_h^2=-M_Z^2$ is motivated by the fact that the vector form factor is evaluated at a time-like momentum transfer. In \cite{Ahrens:2008qu}, it was shown that this choice greatly improves the convergence of the expansion of the Higgs-boson production cross section. In the Drell-Yan case the effect is less pronounced \cite{Ahrens:2008qu,Ahrens:2008nc,Magnea:1990zb}. At NNLL order (NLO in RG-improved perturbation theory), which is the relevant approximation for our paper, the corrections are similar in both schemes, but at N$^3$LL (corresponding to NNLO) the corrections and the associated scale uncertainty become much smaller for $\mu^2=-M_Z^2$. Using the values in the table, one can easily read off the uncertainties associated with the variation of $\mu_h$ and switch between the two schemes. Our NNLL plots are given for $\mu^2=-M_Z^2$, so that the hard scale uncertainty is $(^{+0.1}_{-1.0})\%$. To obtain the result for $\mu^2=+M_Z^2$, one multiplies by $0.961$ and obtains a scale uncertainty of $(^{+1.0}_{-0.1})\%$. In view of the known N$^3$LL result for the hard function, it is likely that the value obtained with $\mu^2=-M_Z^2$ is more reliable and that the scale uncertainties obtained with $\mu^2=M_Z^2$ underestimate the higher-order corrections.

\begin{figure}[t!]
\begin{center}
\begin{tabular}{rcr}
\includegraphics[width=0.45\textwidth]{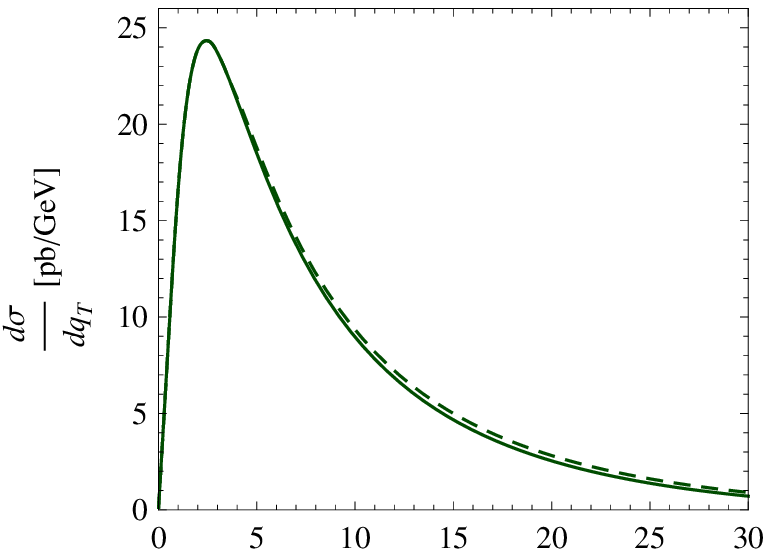} & &
\includegraphics[width=0.45\textwidth]{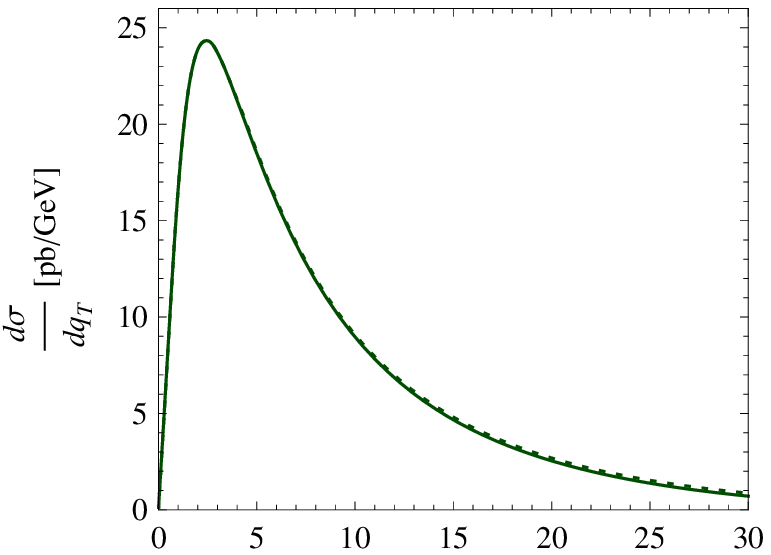} \\[-0.0cm]
\includegraphics[width=0.42\textwidth]{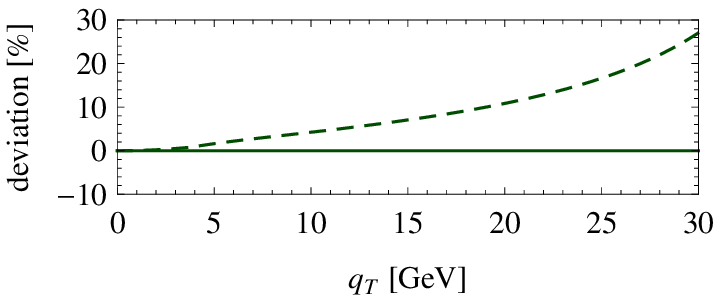} & &
\includegraphics[width=0.42\textwidth]{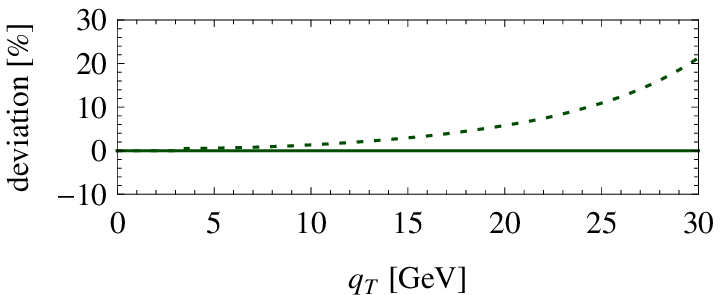} 
\end{tabular}
\end{center}
\vspace{-0.5cm}
\caption{\label{Schemes}
Comparison of different ways to perform the perturbative expansion and scale setting. The solid line in the two plots shows our default prediction, the dashed curve in the left plot is obtained by performing the $\epsilon$ expansion in the exponent, while the dotted curve in the right plot is obtained if the scale setting is performed for the integrated cross section.}
\end{figure}

In Figure~\ref{Schemes}, we compare different ways to perform the perturbative expansion in RG-improved perturbation theory. For example, instead of expanding out the higher-order terms in $\epsilon$ contained in $g_F$ and in the hard function, we can keep these terms in the exponent. The result is shown by the dashed line in the left plot. It is gratifying to see that the difference between expanding out and keeping the terms is indeed very small in the peak region, such that the lines obtained with the two prescriptions are nearly indistinguishable. The small difference also justifies treating the hard function as an overall factor. In the numerical implementation of the traditional CSS formalism, the scale setting is often performed for the integrated rate 
\begin{equation}\label{intsigma}
   \Sigma(q_T) = \int_0^{q_T}\!dq_T'\,\frac{d\sigma}{dq_T'} \,,
\end{equation}
and the spectrum is then obtained as the derivative of $\Sigma(q_T)$. As long as the renormalization scales are held fixed, integrating and differentiating commute. However, if the scale is chosen in a $q_T$-dependent way, $\mu=q_*+q_T$, there is higher-order scale dependence in which the results obtained from $\Sigma(q_T)$ differ from a direct evaluation of the spectrum. An advantage of setting the scale in $\Sigma(q_T)$ is that one automatically recovers the fixed-order result if one integrates the resummed spectrum to high values of $q_T$, since the entire integral is then evaluated at a high scale. One can argue whether this is an important requirement, since after all the resummation was designed for the low $q_T$ region, not for the total cross section. Either way, the right plot in the figure makes it clear that the associated difference is very small. At larger values $q_T\gtrsim 15$\,GeV the differences with respect to our default choice, shown in the panels below the plots, are no longer negligible. To reduce the scheme dependence in this region one will need to match to ${\cal O}(\alpha_s^2)$ fixed-order results, since the bulk of the difference arises from ${\cal O}(\alpha_s^2)$ terms beyond the accuracy of our computation.

\begin{figure}[t!]
\begin{center}
\begin{tabular}{cc}
\includegraphics[width=0.42\textwidth]{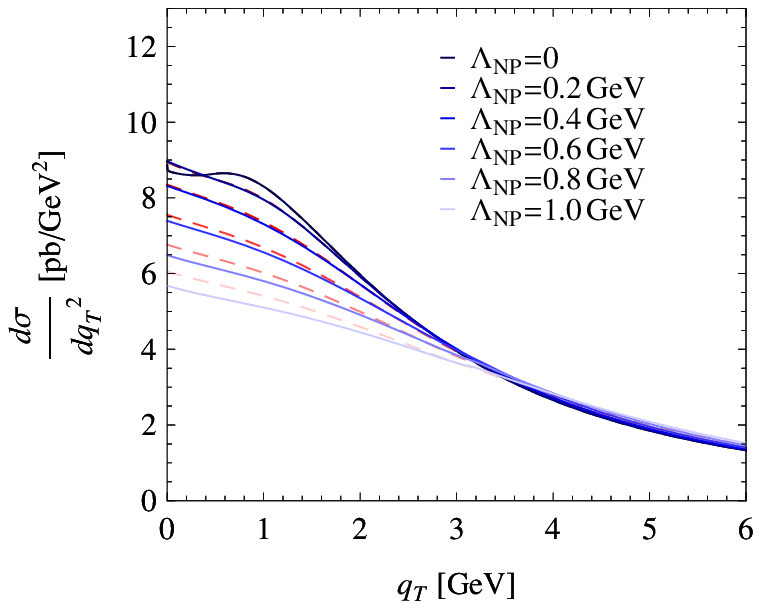} \hspace{1mm} & 
\includegraphics[width=0.42\textwidth]{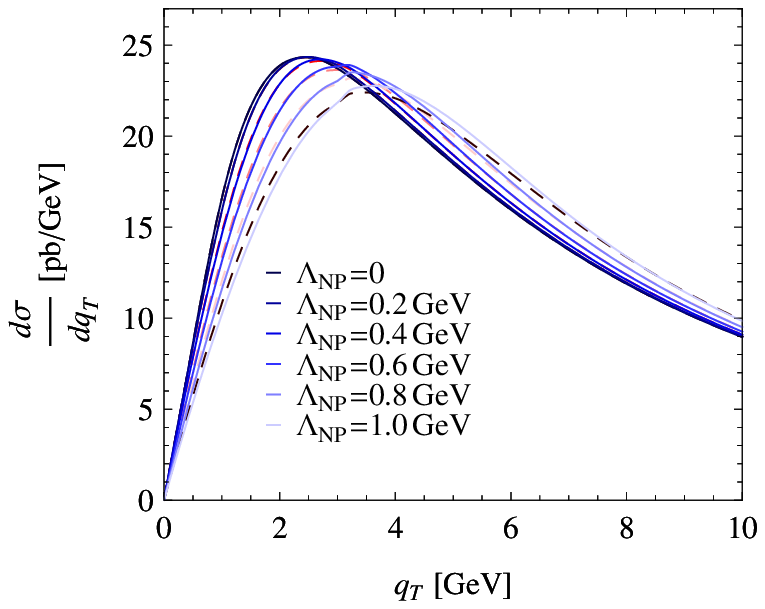} \\
\includegraphics[width=0.42\textwidth]{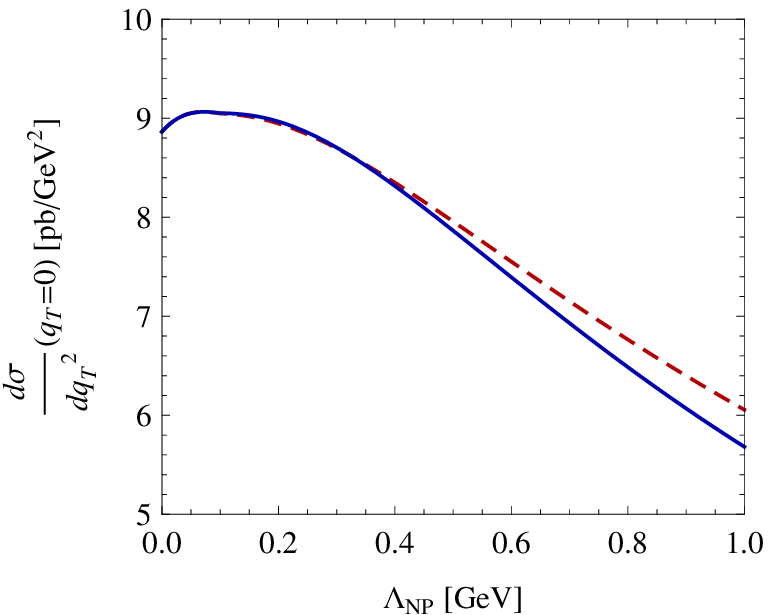} & \hspace{3mm}
\includegraphics[width=0.395\textwidth]{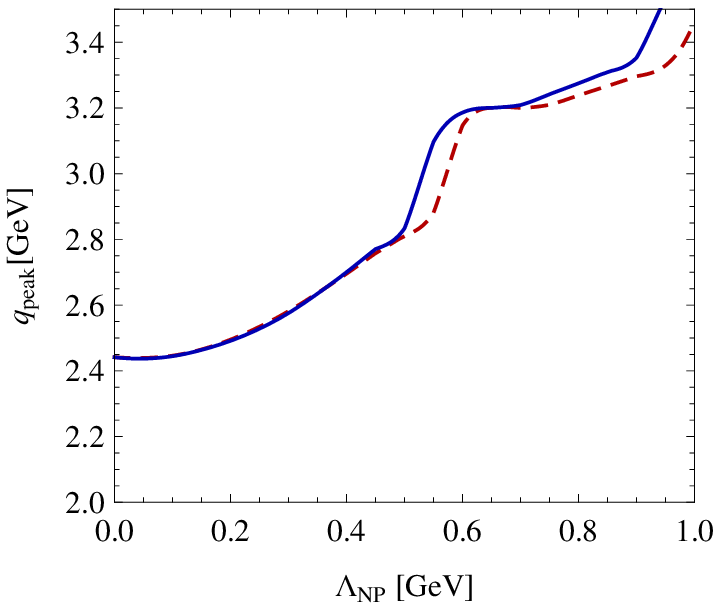}
\end{tabular}
\vspace{-0.5cm}
\end{center}
\caption{\label{nonpert}
Long-distance effects on the differential cross sections (upper row), the intercept of $d\sigma/dq_T^2$ (lower left), and the peak of $d\sigma/dq_T$ (lower right). The solid blue lines are obtained from modeling with a gaussian, while the dashed red lines correspond to a pole form.}
\end{figure}

Before comparing to data, we now study the effect of long-distance corrections. As discussed near the end of Section~\ref{sec:intercept}, because of the violent divergence of the expansion of the cross section around $q_T=0$ it is not possible to perform the usual twist expansion to study these corrections. Instead, we will model long-distance corrections using form factors such as the ones shown in (\ref{fmodels}) under the Fourier integral in (\ref{Cdef}), where $\Lambda_{\rm NP}$ is a typical scale associated with low-energy QCD. These form factors suppress the region of large $x_T$. The effect of the non-perturbative corrections on the spectrum is shown in Figure~\ref{nonpert} for $0\leq\Lambda_{\rm NP}\leq 1$\,GeV. Already for $q_T\geq 3$\,GeV, the effects are almost negligible in $d\sigma/dq_T^2$. It is also remarkable that they are to a large extent insensitive to the shape of the form factor. The dashed red lines obtained with the pole form lie very close to the solid blue lines obtained with the gaussian form. For $d\sigma/dq_T$, the corrections result in a small shift of the distribution. The dependence of this shift on the parameter $\Lambda_{\rm NP}$ is shown in the fourth plot in Figure~\ref{nonpert} and is quite similar for the two different cut-offs. The kink in the peak position arises because the peak happens to move over the $b$-quark flavor threshold as $\Lambda_{\rm NP}$ is increased beyond 500\,MeV. We integrate out the $b$ quark at a scale $\mu_b=4.75$\,GeV, which corresponds to a $q_T$ value of $q_T=\mu_b-q_*\approx 2.9$\,GeV.

\begin{figure}[t!]
\begin{center}
\begin{tabular}{cc}
\includegraphics[height=0.35\textwidth]{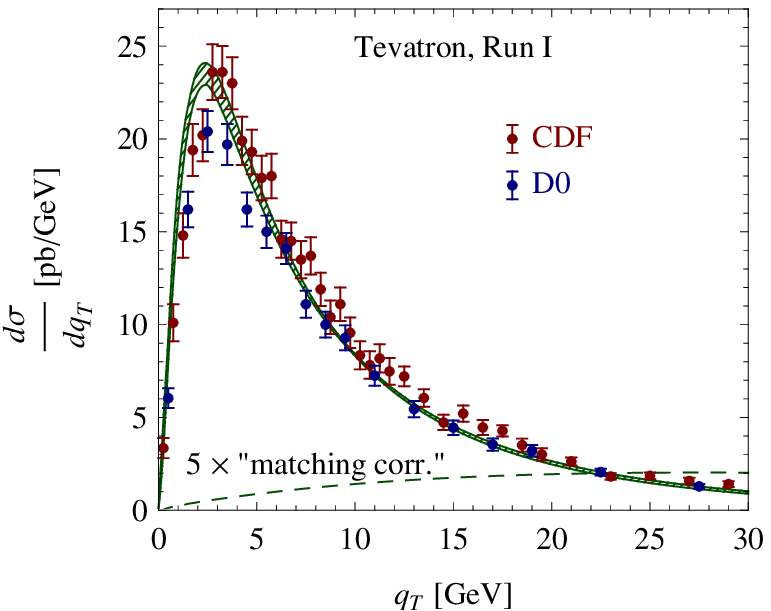} & 
\includegraphics[height=0.35\textwidth]{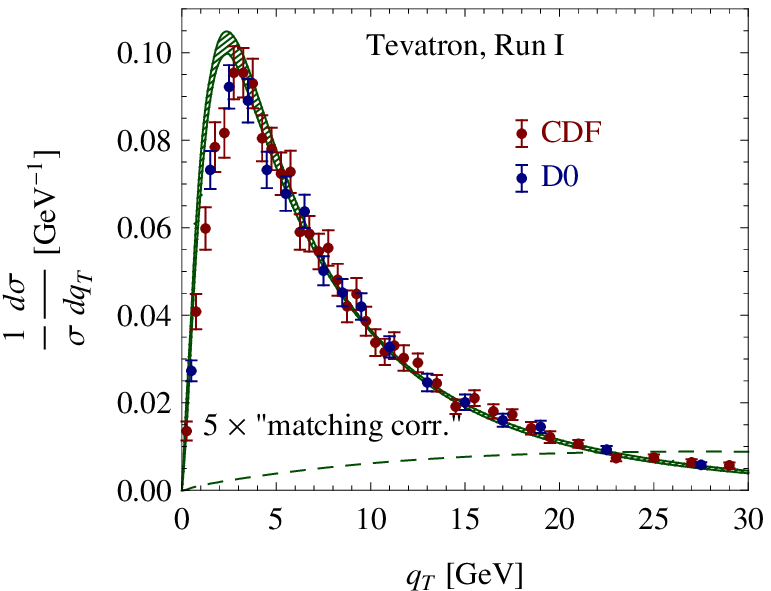}
\end{tabular}
\vspace{-0.5cm}
\end{center}
\caption{\label{runI}
Comparison to Run I data from {\sc CDF} \cite{Affolder:1999jh} and D\O\ \cite{Abbott:1999yd}.}
\end{figure}

\begin{figure}[t!]
\begin{center}
\begin{tabular}{rcr}
\includegraphics[width=0.45\textwidth]{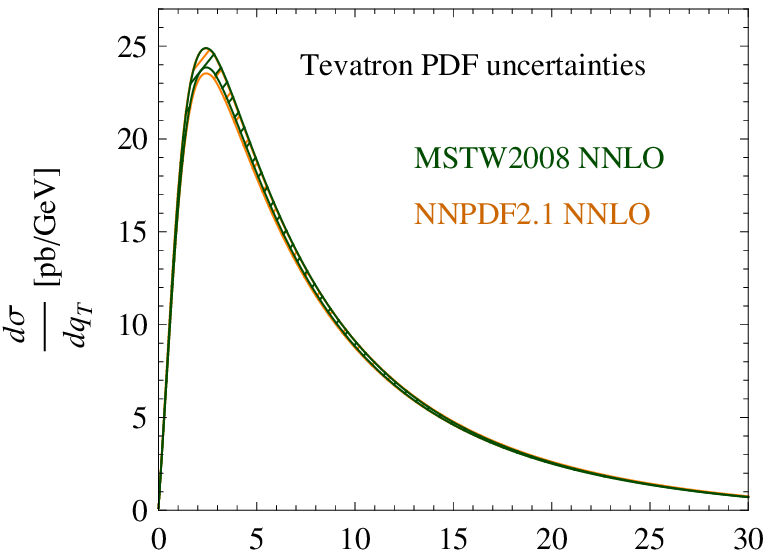} & &
\includegraphics[width=0.45\textwidth]{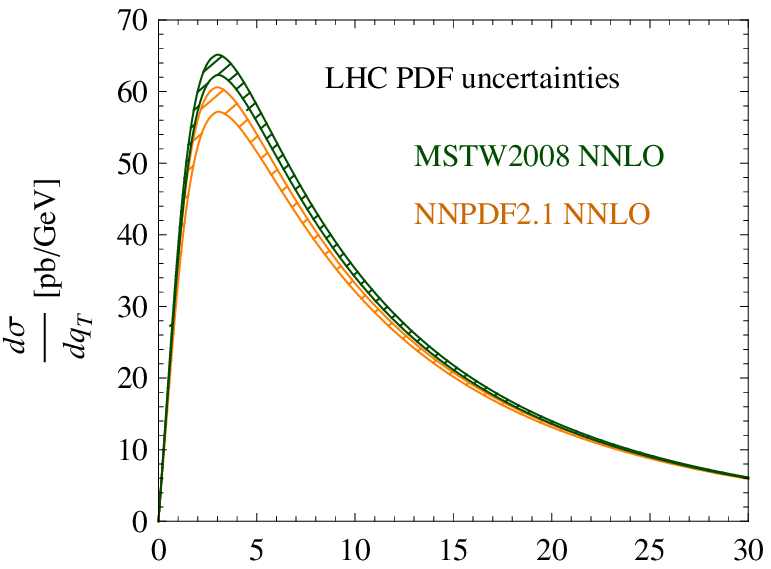} \\[-0.5cm]
\includegraphics[width=0.42\textwidth]{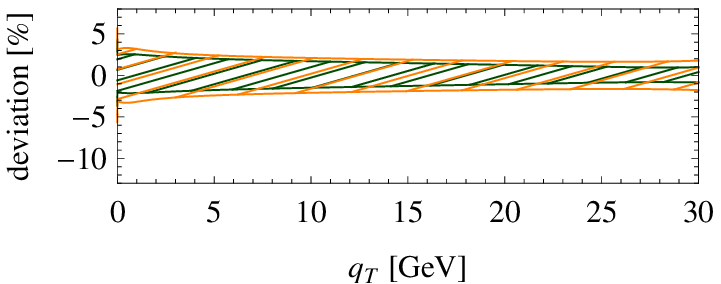} & &
\includegraphics[width=0.42\textwidth]{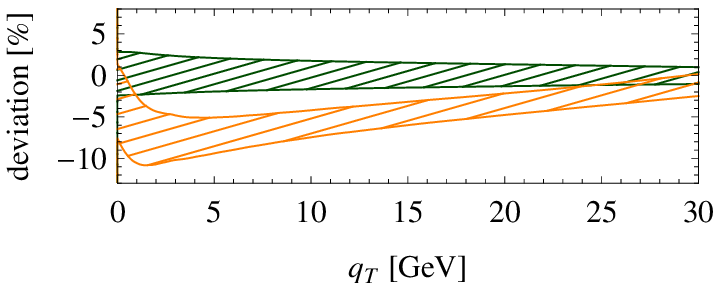} 
\end{tabular}
\end{center}
\vspace{-0.5cm}
\caption{\label{pdfuncertainty}
Study of PDF uncertainties, where the bands correspond to one standard deviation.}
\end{figure}

We have mentioned in Section~\ref{sec:intercept} that we expect the dynamically generated scale $q_*$ to set the size of hadronic power corrections to the short-distance results obtained in this paper. With the example of the dependence of the intercept of the $d\sigma/dq_T^2$ distribution on the hadronic parameter $\Lambda_{\rm NP}$ we have tested this assertion numerically, studying a large range of values of $\Lambda_{\rm NP}$ and the boson mass $M_V$ (and hence $q_*$). We have confirmed that the resulting relative shifts can be well approximated by a power law $\sim\Lambda_{\rm NP}^2/(q_*^\delta\,\Lambda_{\rm QCD}^{2-\delta})$, with $\delta\approx 1.5$ for the Tevatron and $\delta\approx 1.0$ for the LHC with $\sqrt{s}=14$\,TeV.

\section{Comparison to experimental data}
\label{sec:data}

We now compare our results to the available experimental data. The most detailed picture of the low-$q_T$ region is provided by the results by {\sc CDF} \cite{Affolder:1999jh} and D\O\ \cite{Abbott:1999yd} obtained during Run~I of the Tevatron, which are quite finely binned at small transverse momentum. In Figure~\ref{runI}, we plot the the experimental data, together with our prediction, obtained using the improved expansion scheme at NLO, matched to the ${\cal O}(\alpha_s)$ fixed-order result. The small matching corrections are shown by dashed lines in the figure and have been multiplied by a factor 5 to make them visible on the scale of the plots. For the hard scale we choose the time-like value $\mu_h^2=-M_Z^2$, and we use $\mu=q_T+q_*$ for the factorization scale, which is varied by a factor of two to obtain the uncertainty bands in the various plots. We do not show the small uncertainty of $^{+0.1\%}_{-1\%}$ associated with varying the hard scale. It is $q_T$-independent and can be found in Table \ref{tabhard}. The PDF uncertainties are shown in Figure~\ref{pdfuncertainty} and are of similar size at the Tevatron and the LHC. At low $q_T$, they are of order 2.5\% and then decrease linearly to roughly 1\% near $q_T=30\,{\rm GeV}$. In the same figure, we also show the predictions obtained using NNPDF~2.1 PDF sets \cite{Ball:2011mu}. They give somewhat larger uncertainties. The central value is almost identical to MSTW at the Tevatron, but about 7\% lower in the peak region for the LHC.

\begin{figure}[t!]
\begin{center}
\begin{tabular}{rcr}
\includegraphics[width=0.455\textwidth]{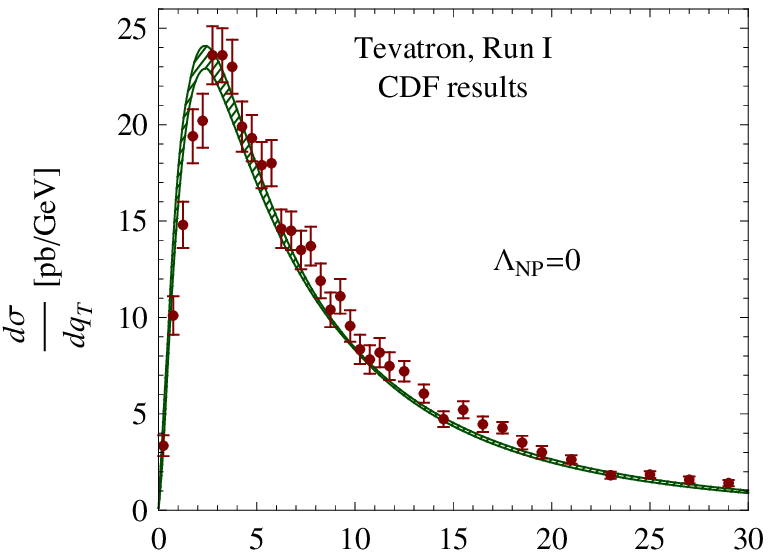} & &
\includegraphics[width=0.455\textwidth]{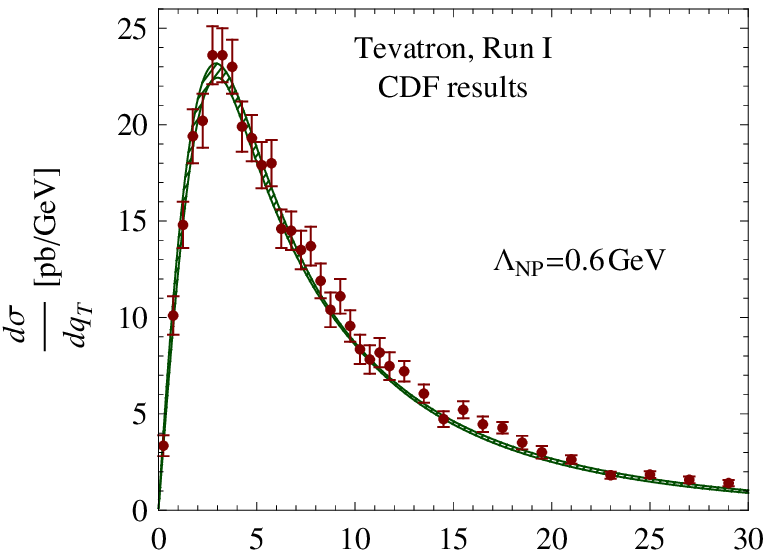} \\[-0.5cm]
\includegraphics[width=0.429\textwidth]{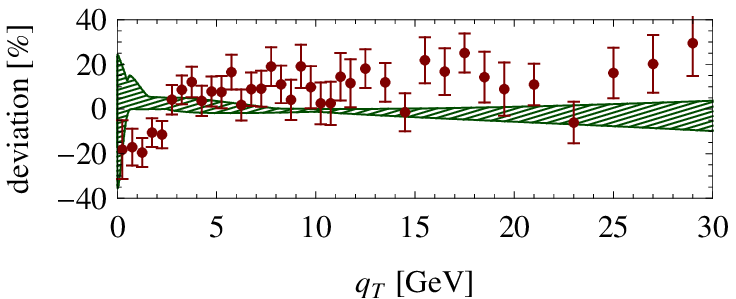} & &
\includegraphics[width=0.429\textwidth]{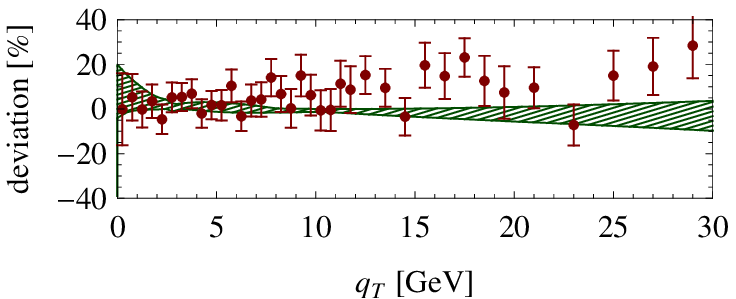} 
\end{tabular}
\end{center}
\vspace{-0.5cm}
\caption{\label{nonpertshift}
Comparison with Tevatron Run I data from {\sc CDF}, with and without long-distance corrections. The lower panels show the deviation from the default theoretical prediction.}
\end{figure}

The overall agreement of the data \cite{Affolder:1999jh,Abbott:1999yd} with our result is very good, but the D\O\ data is consistently lower than the {\sc CDF} result and our theoretical prediction. Summing the data bins to the total cross section  $p\bar{p} \to Z+X \to \ell^+\ell^- +X$, one finds that the result of {\sc CDF} amounts to $\sigma_{\rm tot} =247.4\, {\rm pb}$, while D\O\ obtains $\sigma_{\rm tot} =221.3\,{\rm pb}$. The theoretical prediction for the total cross section at ${\cal O}(\alpha_s^2)$ is $\sigma_{\rm tot} =229.7\,{\rm pb}$, in between the two values.\footnote{We have used the code VRAP \cite{vrap}, based on the paper \cite{Anastasiou:2003ds}, to obtain the total cross section.} On the right hand side, we show the result obtained by normalizing the data to their respective total cross section. One observes that the shape agrees well between the two experiments. We have normalized the theoretical curve to the ${\cal O}(\alpha_s^2)$ fixed order result. Note that this does not guarantee that the theoretical prediction for the spectrum is normalized to one, since we have  only matched to ${\cal O}(\alpha_s)$ in the spectrum. 

\begin{figure}[t!]
\begin{center}
\begin{tabular}{rcr}
\includegraphics[width=0.455\textwidth]{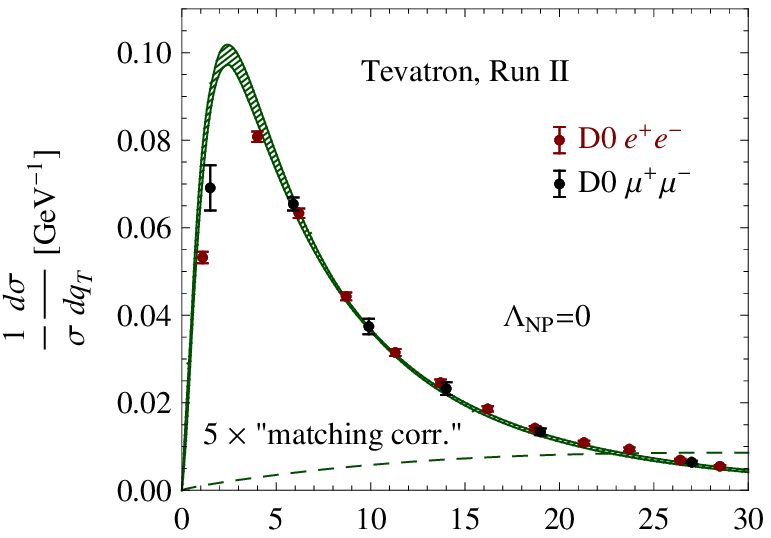} & &
\includegraphics[width=0.455\textwidth]{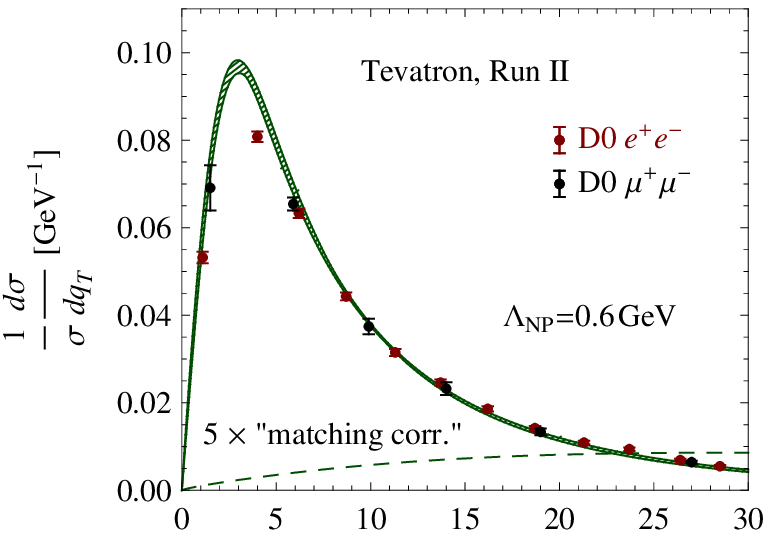} \\[-0.5cm]
\includegraphics[width=0.416\textwidth]{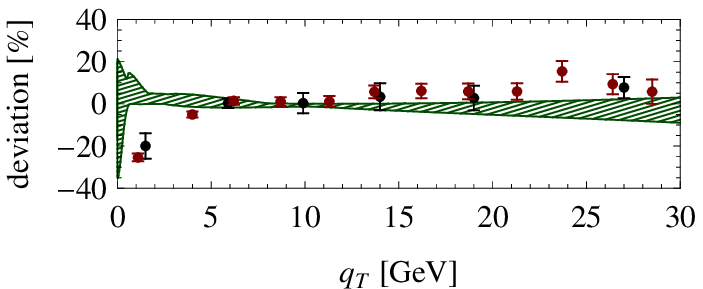} & &
\includegraphics[width=0.416\textwidth]{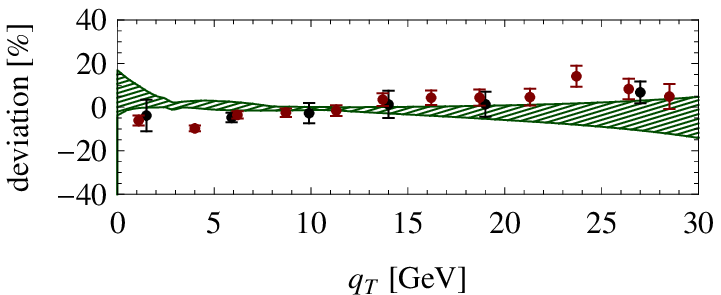} \\[0.5cm]
\includegraphics[width=0.455\textwidth]{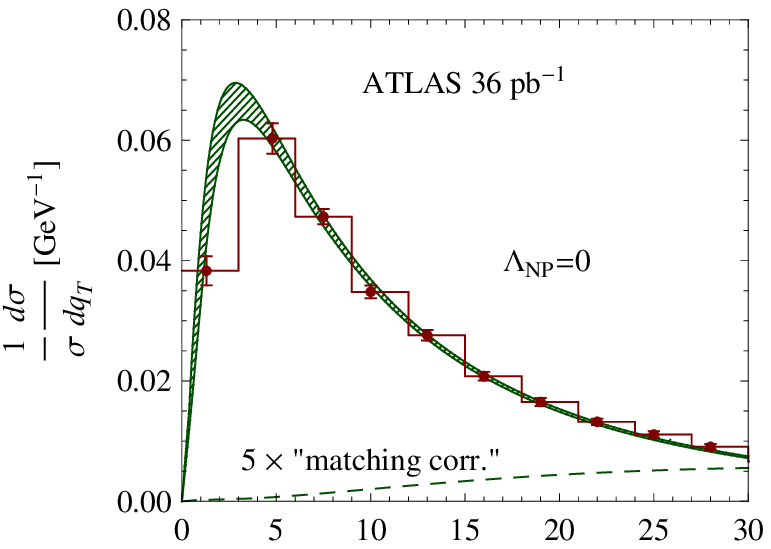} & &
\includegraphics[width=0.455\textwidth]{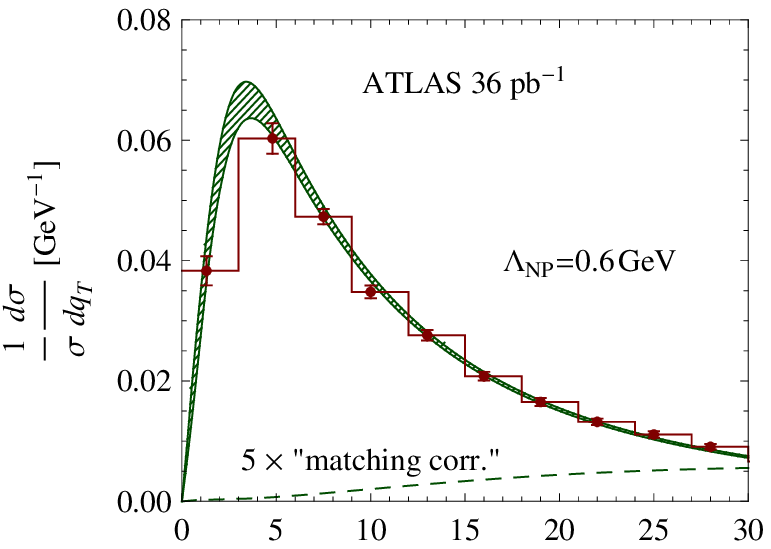} \\[-0.5cm]
\includegraphics[width=0.416\textwidth]{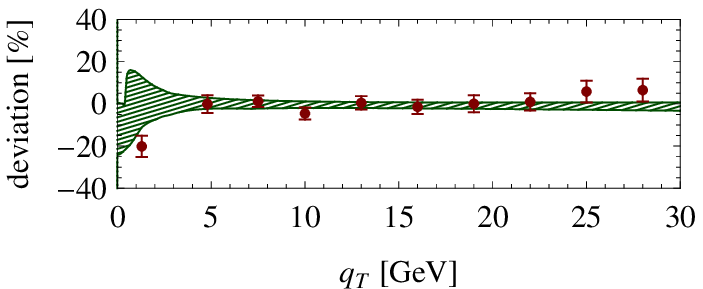} & &
\includegraphics[width=0.416\textwidth]{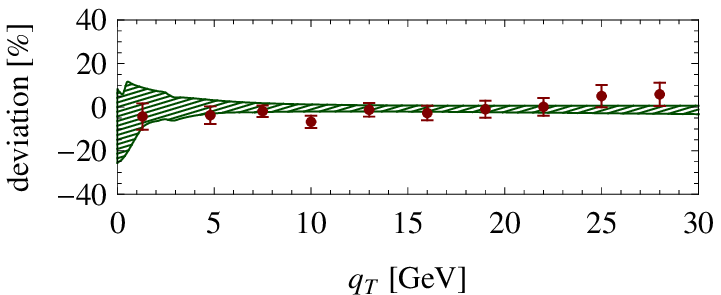} 
\end{tabular}
\end{center}
\vspace{-0.5cm}
\caption{\label{tevandlhc}
Comparison to Tevatron Run II and ATLAS data, with and without long-distance corrections. The lower panels show the deviation from the default theoretical prediction.}
\end{figure}

In the plots in Figure~\ref{runI} the observed peak is slightly (by about $750\,{\rm MeV}$) to the right of the prediction, which was obtained setting the non-perturbative parameter $\Lambda_{\rm NP}=0$ to zero. We have discussed in the previous section that long-distance corrections will shift the peak to the right, and Figure~\ref{nonpert} shows that a shift of $0.75\,{\rm GeV}$ corresponds to a value of $\Lambda_{\rm NP}=0.6\,{\rm GeV}$. In Figure~\ref{nonpertshift}, we compare again to the {\sc CDF} data \cite{Affolder:1999jh} and plot the theoretical prediction for both $\Lambda_{\rm NP}=0$ and $\Lambda_{\rm NP}=0.6\,{\rm GeV}$. In the lower panels, we give the ratio of the experimental and theoretical results to our default prediction. Including a non-perturbative shift, a good description of the data is achieved over the entire $q_T$ range. In Figure \ref{tevandlhc}, we repeat the same comparison for the Tevatron Run II results from D\O\ \cite{:2007nt,Abazov:2010kn}  and for the LHC result of the ATLAS collaboration \cite{Aad:2011gj}. Since this data is not finely binned in the peak region, it difficult to draw firm conclusions on the necessity for long-distance corrections. However, in both cases, the first data bin is below the prediction without including a long-distance correction. 

The systematic experimental uncertainties which affect the low $q_T$ experimental results are substantial, because it is highly sensitive to lepton transverse momentum resolution. Recently, two new variables $a_T$ and $\phi^*_\eta$  were introduced, which probe the same physics but have  reduced sensitivity to the momentum resolution \cite{Vesterinen:2008hx,Banfi:2010cf}. D\O\ has now performed a very precise measurement of the variable  $\phi^*_\eta$ \cite{Abazov:2010mk}. It would be interesting to include the lepton decay in our results and to study these variables. In the traditional framework, resummed results for these quantities were presented recently in \cite{Banfi:2011dx,Marzani:2011yf}.

\begin{figure}[t!]
\begin{center}
\begin{tabular}{rcr}
\includegraphics[width=0.455\textwidth]{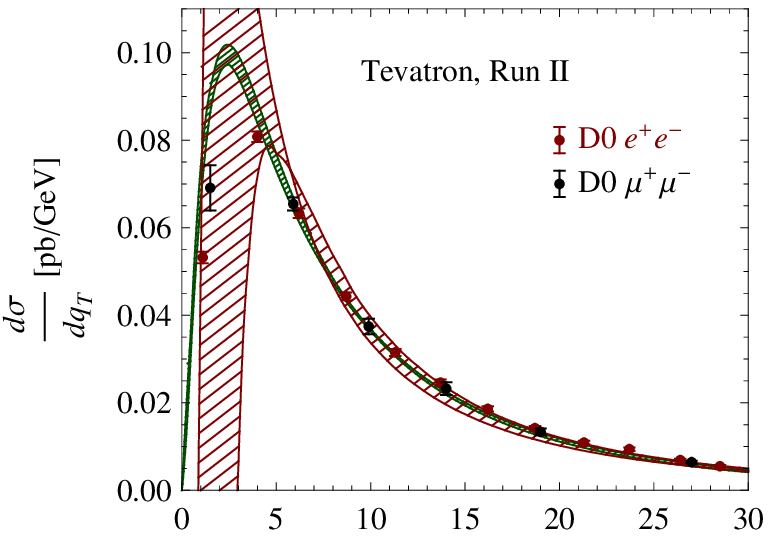} & &
\includegraphics[width=0.455\textwidth]{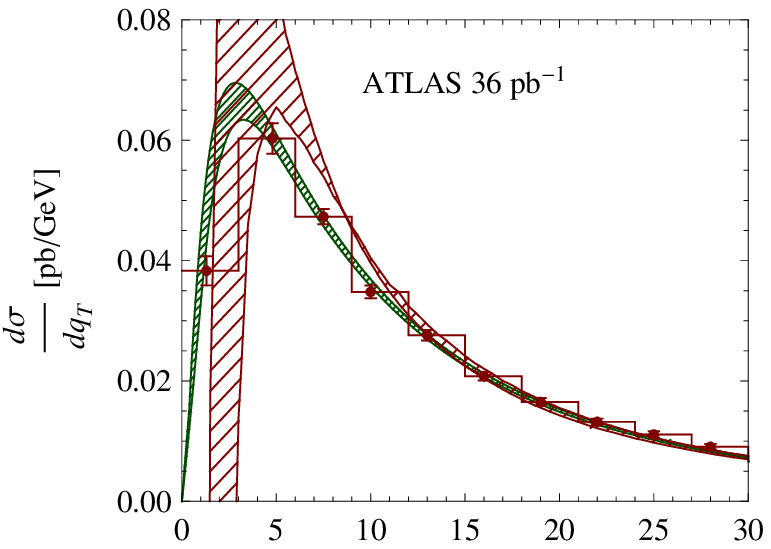} \\[-0.5cm]
\includegraphics[width=0.416\textwidth]{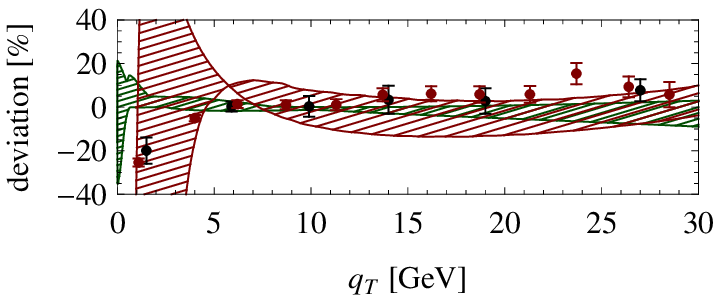} & &
\includegraphics[width=0.416\textwidth]{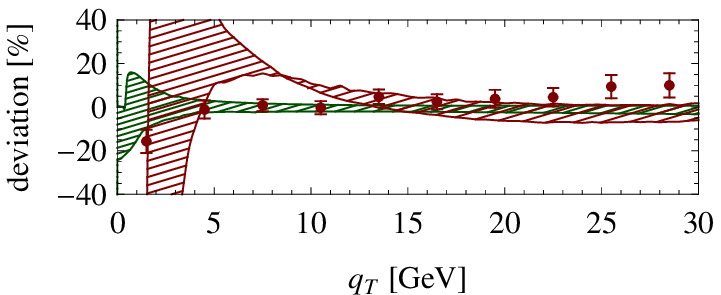} \end{tabular}
\end{center}
\vspace{-0.5cm}
\caption{\label{fixed}
Comparison of the resummed result (green) and ${\cal O}(\alpha_s^2)$ fixed-order result (red) for $q_T$ spectrum at the Tevatron and at LHC.}
\end{figure}

The region of larger $q_T\gtrsim 20\,{\rm GeV}$ is not affected by long-distance corrections and should be described well by fixed-order perturbation theory. In this region the data lies somewhat above the prediction, in particular for the case of the ATLAS results. A comparison to the existing fixed-order results is given in Figure~\ref{fixed}. The red bands correspond to the ${\cal O}(\alpha_s^2)$ fixed-order result for the spectrum, which the highest order currently known. To compute this result we use the numerical code QT \cite{qt}. For the sake of comparison, we have evaluated all results using the MRST2008NNLO PDF set. The fixed-order results diverge to $\pm\infty$ for vanishing transverse momentum. Since the fixed-order result depends both on $q_T$ and $M_Z$ it is not clear which value one should choose for the renormalization and factorization scales. The edges of the fixed-order band in Figure~\ref{fixed} correspond to the two choices $\mu=q_T$ and $\mu=M_Z$. The choice $\mu=q_T$ diverges to $-\infty$, while $\mu=M_Z$ rises to $+\infty$. Our result is consistent with the ${\cal O}(\alpha_s^2)$ fixed-order result at large $q_T$, but performing the matching to this order would presumably lead to a slight increase of our cross-section prediction in this region.

We have also compared our findings to the results \cite{Bozzi:2010xn}. They perform resummation at the same level of accuracy but with a  different formalism, based on the CSS formula \cite{Collins:1984kg}. The peak position is exactly the same as in our result, and their peak height is about 2.5\% lower, but compatible within uncertainties. Since we have adopted a time-like hard matching scale, which gives a 4\% higher cross-section than the usual space-like choice, this difference is not unexpected. Also at  $q_T=30\,{\rm GeV}$, the two results are compatible, but their central value is about 8\% higher than ours, perhaps because the NLO matching has been implemented, or because of the unitarization prescription implemented in their formula. By inspecting plots showing a comparison of results of the RESBOS code \cite{Balazs:1997xd} with Tevatron \cite{Landry:2002ix} and LHC \cite{Aad:2011gj} data, we conclude that our results are compatible with these predictions, which also have N$^2$LL accuracy. We also note that our results have much smaller scale uncertainties than the numerical results obtained in \cite{Mantry:2010mk,Mantry:2011xj} based on the formalism presented in \cite{Mantry:2009qz}. Figure~3 in \cite{Mantry:2011xj} shows that the peak height in $d\sigma/dq_T$ changes roughly by a factor 3 when the factorization scale $\mu$ is varied in the narrow range $2.8\,{\rm GeV}<\mu<4.4\,{\rm GeV}$. In our case, the peak height changes only by 5\%, even though the scale is varied over the larger range $2.1\,{\rm GeV}<\mu<8.6\,{\rm GeV}$. We believe that this numerical instability illustrates our earlier point, namely that the results of \cite{Mantry:2009qz,Mantry:2010mk,Mantry:2011xj} only have LL accuracy (in the exponent), since they do not resum the corrections associated with the collinear anomaly.

\section{Conclusions}
\label{sec:concl}

The transverse momentum spectrum in Drell-Yan processes is one of the most basic observables at hadron colliders. It neverthess manifests a number of remarkable properties at low transverse momentum. These are associated with the collinear anomaly, which complicates the factorization of the cross section. Because of the anomaly, not only the hard function, which encodes the virtual effects associated with the electroweak boson production, but also the product of collinear functions has a dependence on the boson mass $M_V$. In a previous paper, we have shown that the anomalous $M_V$ dependence exponentiates in transverse position space. This guarantees that the associated large logarithms are under control, but at the same time renders the Fourier integral non-trivial, which yields the transverse momentum spectrum. One of the interesting consequences of the presence of the anomaly is that the scale $\mu$, which controls the perturbative corrections to the factorization theorem is no longer $\mu\approx q_T$ at very low $q_T$, but saturates to a non-perturbative value $q^*\sim M_V e^{-{\rm const}/\alpha_s(M_V)}$ for $q_T\to 0$. Numerically, $q^*\approx 1.88\,{\rm GeV}$ for $Z$-production, which implies that the spectrum remains short-distance dominated even in the limit $q_T\to 0$. The fact, that the zero intercept of $d\sigma/dq_T^2$ is calculable for very large Drell-Yan masses was pointed out a long time ago by by Parisi and Petronzio, but our formalism has allowed us to systematically evaluate the corrections to their asymptotic result.  The uncertainty on the calculable short-distance corrections, as well as the long-distance effects, which we estimate using models, are seizable given that the relevant scale is $q^*\approx 1.88\,{\rm GeV}$. Nevertheless, it is possible to obtain a prediction for the intercept for $Z$-production with an accuracy of about 25\%. This demonstrates that, contrary to common belief, the mechanism identified by Parisi and Petronzio is not merely an interesting theoretical result, but of phenomenological relevance.  

In addition to making the spectrum infrared safe, the anomaly precludes the expansion of the spectrum in any of its parameters. In particular, there is a strong factorial divergence in the expansion in $\alpha_s$, associated with terms in the Fourier integral which become enhanced near $q_T=0$. While they are not logarithmically enhanced, these terms nevertheless need to be resummed. An even more violent divergence arises when one tries to expand the spectrum around $q_T^2=0$. In this case the $n$-th order terms in the expansion are enhanced by $e^{n^2}$, which renders the expanded result meaningless. To study long-distance effects, we  insert a cut-off function such as a Gaussian $\exp(-\Lambda_{\rm NP}^2 x_T^2 )$ into the Fourier integral.  We find that the long-distance effects are of the expected size and largely independent of the form of the cut-off. However, a twist expansion $\exp(-\Lambda_{\rm NP}^2 x_T^2 )=1-\Lambda_{\rm NP}^2 x_T^2+ \dots$ suffers from the  same strong divergence as the expanision around $q_T=0$ and can thus not be used.

We have compared to experimental results from both runs of the Tevatron and from the LHC. We find excellent agreement with our predictions, however, for $q_T \lesssim 3\, {\rm GeV}$ the agreement is only achieved after including long-distance effects.  The same choice of the associated parameter $\Lambda_{\rm NP} = 0.6\,{\rm GeV}$ describes these effects both at the Tevatron and the LHC, which provides a consistency check on the model we used. It would be interesting to study the long-distance effects in more detail. To this end, the recently proposed variables $a_T$ and $\phi_\eta^*$ would be especially well suited, and the variable  $\phi_\eta^*$ has now been measured by DZero. To compute these quantities, one needs to include the decay of the electroweak boson  into the lepton pair. It is straightforward to extend our results to this case, and it is important in order to be able to implement the lepton cuts used by the experiments, and to compute the charged-lepton spectrum used for the $W$-mass determination. 

The formalism developed here can be applied to other Drell-Yan-type processes, such as Higgs-boson production or the production of new heavy, color-neutral particles at hadron colliders. The case of Higgs production via gluon-gluon fusion is particularly interesting. In this case many of our formulas apply with a simple substitution of color factors. In particular, in the expressions for the dynamically generated scale $q_*$ in (\ref{qstar}) and for the exponent $g_F$ of the Bessel integral in (\ref{gFmod}), one must replace the one-loop cusp anomalous dimension in the fudamental representation with that in the adjoint representation, $\Gamma_0^F\to\Gamma_0^A$. As a result, for a Higgs with mass of 120\,GeV, for example, the value of $q_*$ is 7.5\,GeV. This is truly a short-distance scale, and all of the methods developed here should work much more accurately than in the case of $Z$-boson production. In particular, long-distance effects in the region of very small $q_T$ should be strongly suppressed. A phenomenological analysis of Higgs production in our framework is left for future work.

\vspace{0.2cm}
{\em Acknowledgments:\/}
Part of this research was performed at the KITP Santa Barbara, while two of us (T.B.\ and M.N.) were attending the program {\em The Harmony of Scattering Amplitudes}. We are grateful to the KITP for the hospitality and support. The research of M.N.\ and D.W.\ is supported in part by BMBF grant 05H09UME, DFG grant NE 398/3-1, and the Research Centre {\em Elementary Forces and Mathematical Foundations\/}. T.B.\ is supported in part by the SNSF and the ``Innovations- und Kooperationsprojekt C-13'' of SUK.

\vspace{2cm}
\begin{appendix}

\section{Single differential cross section}
\label{app:a}
\renewcommand{\theequation}{A\arabic{equation}}
\setcounter{equation}{0}

Integrating the double differential cross section (\ref{fact1}) over rapidity, we obtain
\begin{equation}
\begin{aligned}
   \frac{d\sigma}{dq_T^2} 
   &= \frac{4\pi^2\alpha}{N_c\,s} \left| C_V(-M_Z^2,\mu) \right|^2
    \sum_q\,\frac{|g_L^q|^2+|g_R^q|^2}{2}\,\sum_{i=q,g} \sum_{j=\bar q,g} 
    \int_{\tau}^1\!\frac{dz}{z} \\
   &\quad\times \bigg[ \tilde C_{q\bar q\leftarrow ij}\big(z,q_T^2,M_Z^2,\mu\big)\,
    f\hspace{-1.8mm}f_{ij}(\tau/z,\mu) + (q,i\leftrightarrow\bar q,j) \bigg] \,,
\end{aligned}
\end{equation}
where $g_L^q=(g/e)\,(T_3^q-e_q\sin^2\theta_W)$ and $g_R^q=(g/e)\,(-e_q\sin^2\theta_W)$ denote the couplings (in units of $e$) of the $Z$ boson to left-handed and right-handed quarks, and hence 
\begin{equation}
   \frac{|g_L^q|^2+|g_R^q|^2}{2}
   = \frac{\big( 1 - 2|e_q|\sin^2\theta_W \big)^2 + 4 e_q^2\sin^4\theta_W}%
          {8\sin^2\theta_W\cos^2\theta_W} \,.
\end{equation}
The parton luminosities are defined as
\begin{equation}
   f\hspace{-1.8mm}f_{ij}(z,\mu)
   = \int_z^1\!\frac{du}{u}\,\phi_{i/N_1}(u,\mu)\,\phi_{j/N_2}(z/u,\mu) 
   \equiv \Big( \phi_{i/N_1}(\mu)\otimes\phi_{j/N_2}(\mu) \Big)(z) \,.
\end{equation}
The kernel functions $\tilde C_{q\bar q\leftarrow ij}$ are obtained from the functions $\bar C_{q\bar q\leftarrow ij}$ defined in (\ref{Cdef}) by a convolution in the two $z_i$ variables, such that
\begin{equation}
\begin{aligned}
   \tilde C_{q\bar q\leftarrow ij}(z,q_T^2,M_Z^2,\mu)
   &= \frac{1}{4\pi} \int\!d^2x_\perp\,e^{-iq_\perp\cdot x_\perp}
    \left( \frac{x_T^2 M_Z^2}{b_0^2} \right)^{-F_{q\bar q}(L_\perp,a_s)} \\
   &\quad\times \Big( I_{q\leftarrow i}(L_\perp,a_s)\otimes
    I_{\bar q\leftarrow j}(L_\perp,a_s) \Big)(z) \,.
\end{aligned}
\end{equation}

\section{Hard matching coefficient}
\label{app:hard}
\renewcommand{\theequation}{B\arabic{equation}}
\setcounter{equation}{0}

The solution to the RG equation (\ref{Cevol}) for the hard function takes the general form \cite{Becher:2006mr}
\begin{equation}\label{CVsol}
   \left| C_V(-M_Z^2,\mu) \right|^2 
   = \exp\left[ 4S(\mu_h,\mu) - 4a_{\gamma^q}(\mu_h,\mu) \right]
    \left( \frac{M_Z^2}{\mu_h^2} \right)^{-2a_\Gamma(\mu_h,\mu)}\,
    \left| C_V(-M_Z^2,\mu_h) \right|^2 ,
\end{equation}
where $\mu_h^2\sim-M_Z^2$ is a hard matching scale, at which the value of $C_V$ is calculated using fixed-order perturbation theory. At one-loop order
\begin{equation}\label{CV2fopt}
   \left| C_V(-M_Z^2,\mu_h) \right|^2 = 1 + \frac{C_F\alpha_s(\mu_h)}{2\pi}\,
   \mbox{Re} \left( - L^2 + 3L - 8 + \frac{\pi^2}{6} \right) + \dots \,,
\end{equation}
where $L=\ln(-M_Z^2/\mu_h^2)$. The two-loop correction can also be found in \cite{Becher:2006mr}. The advantages of using a time-like scale choice ($\mu_h^2<0$) for time-like processes such as Drell-Yan production were emphasized in \cite{Ahrens:2008qu,Ahrens:2008nc}. The Sudakov exponent $S$ and the exponents $a_n$ are given by \cite{Neubert:2004dd}
\begin{equation}\label{RGEsols}
\begin{aligned}
   S(\mu_h,\mu) 
   &= \frac{\Gamma_0^F}{4\beta_0^2}\,\Bigg\{
    \frac{4\pi}{\alpha_s(\mu_h)} \left( 1 - \frac{1}{r} - \ln r \right)
    + \left( \frac{\Gamma_1^F}{\Gamma_0^F} - \frac{\beta_1}{\beta_0}
    \right) (1-r+\ln r) + \frac{\beta_1}{2\beta_0} \ln^2 r \\
   &\hspace{1.6cm}\mbox{}+ \frac{\alpha_s(\mu_h)}{4\pi} \Bigg[ 
    \left( \frac{\Gamma_1^F\beta_1}{\Gamma_0^F\beta_0} 
    - \frac{\beta_2}{\beta_0} \right) (1-r+r\ln r)
    + \left( \frac{\beta_1^2}{\beta_0^2} 
    - \frac{\beta_2}{\beta_0} \right) (1-r)\ln r \\
   &\hspace{3.7cm}
    \mbox{}- \left( \frac{\beta_1^2}{\beta_0^2} 
    - \frac{\beta_2}{\beta_0}
    - \frac{\Gamma_1^F\beta_1}{\Gamma_0^F\beta_0} 
    + \frac{\Gamma_2^F}{\Gamma_0^F} \right) \frac{(1-r)^2}{2} \Bigg] + \dots \Bigg\} \,, \\
   a_\Gamma(\mu_h,\mu) 
   &   = \frac{\Gamma_0^F}{2\beta_0} \left[ \,\ln\frac{\alpha_s(\mu)}{\alpha_s(\mu_h)}
    + \left( \frac{\Gamma_1^F}{\Gamma_0^F} - \frac{\beta_1}{\beta_0} 
    \right) \frac{\alpha_s(\mu) - \alpha_s(\mu_h)}{4\pi} + \dots \right] , 
\end{aligned}
\end{equation}
where $r=\alpha_s(\mu)/\alpha_s(\mu_h)$. A similar expression, with the coefficients $\Gamma_i^F$ replaced by $\gamma_i^q$, holds for the function $a_{\gamma^q}$. The relevant expansion coefficients of the anomalous dimensions and $\beta$-function can be found, e.g., in \cite{Becher:2006mr}.

For the special scale choices $\mu_h=M_Z$ and $\mu=q_*$, the expression for the hard matching coefficient can be simplified, since the two couplings $\alpha_s(M_Z)$ and $\alpha_s(q_*)$ are related via the condition that $\eta=1$. Solving the relation
\begin{equation}
   \int\limits_{\alpha_s(q_*)}^{\alpha_s(M_Z)}\!\frac{d\alpha}{\beta(\alpha)}
   = \frac12\,\ln\frac{M_Z^2}{q_*^2} = \frac{1}{2\Gamma_0^F a_s}
\end{equation}
with $a_s\equiv\alpha_s(q_*)/(4\pi)$ iteratively, we derive
\begin{equation}
   \frac{\alpha_s(q_*)}{\alpha_s(M_Z)}
   = \frac{1+c}{c} + a_s\,\frac{\beta_1}{\beta_0}\,\ln\frac{1+c}{c}  
    + \frac{a_s^2}{1+c} \left[ \frac{\beta_2}{\beta_0} - \frac{\beta_1^2}{\beta_0^2} 
    \left( 1 - c\ln\frac{1+c}{c} \right) \right] + {\cal O}\big[ a_s^3(\mu) \big] \,,
\end{equation}
where we have introduced $c=\Gamma_0^F/\beta_0=16/25$. For the purposes of our discussion in this appendix we neglect the presence of flavor thresholds and use $n_f=4$ active quark flavors throughout. Using the above
relation, we find that
\begin{equation}
   \left| C_V(-M_Z^2,q_*) \right|^2 
   = \exp\left[ \frac{1}{\beta_0 a_s} \left( 1 - (1+c)\ln\frac{1+c}{c} \right) \right]\,
    {\cal N}_F\,\Big[ 1 + 1.51591\,a_s + {\cal O}(a_s^2) \Big] \,,
\end{equation}
where
\begin{equation}
   {\cal N}_F = \exp\left[ - \frac{2\gamma_0^q}{\beta_0}\,\ln\frac{1+c}{c}
    - \frac{\beta_1}{2\beta_0^2}\,c\ln^2\frac{1+c}{c} 
    - \frac{1}{\beta_0} \left( \frac{\Gamma_1^F}{\Gamma_0^F} - \frac{\beta_1}{\beta_0} 
    \right) \left( 1 - c\ln\frac{1+c}{c} \right) \right] 
   = 1.83152 \,.
\end{equation}
The explicit expression for the ${\cal O}(a_s)$ terms is more lengthy, and we refrain from giving it here. 

Combining the above result for the hard matching coefficient with expression (\ref{Cq0fin}) for the collinear kernel functions at $q_T=0$, we obtain for the default choice $\mu=q_*$
\begin{equation}\label{beauty}
\begin{aligned}
   & \left| C_V(-M_Z^2,q_*) \right|^2 \bar C_{q\bar q\leftarrow ij}(z_1,z_2,0,M_Z^2,q_*) \\
   &= \frac{e^{-2\gamma_E}}{q_*^2}\,
    \exp\left\{ \frac{1}{\beta_0 a_s} \left[ 1 - \left( 1+\frac{\Gamma_0^F}{\beta_0} \right)
    \ln\left( 1 + \frac{\beta_0}{\Gamma_0^F}\right) \right] \right\}
    \sqrt{\frac{2\pi}{\left( \Gamma_0^F+\beta_0 \right) a_s}} \\
   &\quad\times {\cal N}_F\,\Big\{ \big( 1 + 3.00445\,a_s \big)\,
    \delta(1-z_1)\,\delta(1-z_2)\,\delta_{qi}\,\delta_{\bar qj} + \dots \Big\} \,,
\end{aligned}
\end{equation}
where the remaining terms in the parenthesis have the same form as in (\ref{Cq0fin}), evaluated at $\mu=q_*$. Note that the non-perturbative exponential factor in the first line can be expressed as $e^{\frac{1}{\beta_0 a_s}\,[\dots]}=\left(q_*^2/M_Z^2\right)^{0.3477}$. Along with (\ref{fact1}), formula (\ref{beauty}) provides an explicit expression for the intercept of the Drell-Yan transverse-momentum spectrum in terms of the scale $q_*$. By using the one-loop expression for the running coupling $\alpha_s(\mu)$ to eliminate $q_*$ in favor of the ratio $M_Z/\Lambda_{\rm QCD}$, we can relate our result to approximate saddle-point expressions for the intercept derived in \cite{Parisi:1979se,Collins:1984kg,Ellis:1997ii}. We find that (\ref{beauty}) is consistent with relation (58) in \cite{Ellis:1997ii} up to subleading terms involving the two-loop $\beta$-function. However, the correct normalization factor ${\cal N}_F$ is derived here for the first time.

\section{Perturbation theory with modified power counting}
\label{app:b}
\renewcommand{\theequation}{C\arabic{equation}}
\setcounter{equation}{0}

For a consistent evaluation of the cross section including ${\cal O}(\epsilon)$ corrections in our modified power-counting scheme, where $L_\perp\sim 1/\sqrt{a_s}$, we need to retain some terms of up to four-loop order in $F_{q\bar q}$, up to three-loop order in $h_F$, and up to two-loop order in the kernels $\bar I_{q\leftarrow i}$. The relevant terms are, however, determined by the RG in terms of known anomalous dimensions and $\beta$-function coefficients. We obtain them by solving the evolution equations (\ref{Fevol}) and (\ref{hevol}) recursively, using that $d/d\ln\mu=2\partial/\partial L_\perp+\beta(\alpha_s)\,\partial/\partial\alpha_s$ \cite{Becher:2010tm}. The terms that need to be retained for the purposes of our work read
\begin{eqnarray}\label{Fhexpansions}
   F_{q\bar q}(L_\perp,a_s)
   &=& a_s\,\Gamma_0^F L_\perp
    + a_s^2 \left[ \Gamma_0^F\beta_0\,\frac{L_\perp^2}{2} + \Gamma_1^F\,L_\perp + d_2^q \right] 
    \nonumber\\
   &&\mbox{}+ a_s^3 \left[ \Gamma_0^F\beta_0^2\,\frac{L_\perp^3}{3}
    + \left( \Gamma_0^F\beta_1 + 2\Gamma_1^F\beta_0 \right) \frac{L_\perp^2}{2} + \dots \right]
    + a_s^4 \left[ \Gamma_0^F\beta_0^3\,\frac{L_\perp^4}{4} + \dots \right] + \dots \,, 
   \nonumber\\
   h_F(L_\perp,a_s)
   &=& a_s \left[ \Gamma_0^F\,\frac{L_\perp^2}{4} - \gamma_0^q\,L_\perp \right]
    + a_s^2 \left[ \Gamma_0^F\beta_0\,\frac{L_\perp^3}{12}
    + \left( \Gamma_1^F - 2\gamma_0^q\beta_0 \right) \frac{L_\perp^2}{4} + \dots \right] 
    \nonumber\\
   &&\mbox{}+ a_s^3 \left[ \Gamma_0^F\beta_0^2\,\frac{L_\perp^4}{24} + \dots \right] + \dots \,, \\
   \bar I_{q\leftarrow i}(z,L_\perp,a_s) 
   &=& \delta(1-z)\,\delta_{qi} + a_s \left[ - {\cal P}_{q\leftarrow i}^{(1)}(z)\,\frac{L_\perp}{2}
     + {\cal R}_{q\leftarrow i}(z) \right] \nonumber\\
   &&\mbox{}+ a_s^2 \left[ \left( {\cal D}_{q\leftarrow i}(z)
    - 2\beta_0\,{\cal P}_{q\leftarrow i}^{(1)}(z) \right) \frac{L_\perp^2}{8} + \dots \right] 
    + \dots \,. \nonumber
\end{eqnarray}
The expressions for $\bar I_{q\leftarrow i}$ involve the convolutions ${\cal D}_{q\leftarrow i}$ of DGLAP splitting functions defined in (\ref{Ddef}). Using the one-loop splitting functions in (\ref{APkernels}) along with
\begin{equation}
\begin{aligned}
   {\cal P}_{g\leftarrow q}^{(1)}(z) 
   &= 4C_F\,\frac{ 1+(1-z)^2 }{z} \,, \\
   {\cal P}_{g\leftarrow g}^{(1)}(z) 
   &= 8C_A \left[ \frac{z}{(1-z)_+} + \frac{1-z}{z} + z(1-z) \right] + 2\beta_0\,\delta(1-z) \,,
\end{aligned}
\end{equation}
we obtain
\begin{eqnarray}
   {\cal D}_{q\leftarrow q}(z) 
   &=& 16 C_F^2\,\Bigg[ 4 \left( \frac{\ln\frac{(1-z)^2}{z}}{1-z} \right)_+ 
    + 3 \left( \frac{1+z^2}{1-z} \right)_+ - 4(1+z) \ln(1-z) + 3(1+z) \ln z \nonumber\\
   &&\hspace{12mm}\mbox{}- 2(1-z) - \frac94\,\delta(1-z) \Bigg] \nonumber\\
   &&\mbox{}+ 16 C_F T_F \left[ \frac{4}{3z} + 1 - z - \frac{4z^2}{3} + 2(1+z) \ln z \right] , \\
   {\cal D}_{q\leftarrow g}(z) 
   &=& 16 C_F T_F \left[ \left( z^2 + (1-z)^2 \right) \ln\frac{(1-z)^2}{z} - 2z^2\ln z
    - \frac12 + 2z \right] \nonumber\\
   &&\mbox{}+ 32 C_A T_F \left[ \left( z^2 + (1-z)^2 \right) \ln(1-z) + (1+4z) \ln z
    + \frac{2}{3z} + \frac12 + 4z - \frac{31z^2}{6} \right] \nonumber\\
   &&\mbox{}+ 8\beta_0 T_F \left[ z^2 + (1-z)^2 \right] . \nonumber
\end{eqnarray}
In deriving the first result, we have used that
\begin{equation}
   \frac{1}{(1-z)_+}\otimes\frac{1}{(1-z)_+} = \left[ \frac{\ln\frac{(1-z)^2}{z}}{1-z} \right]_+ .
\end{equation}

\section{Matching to fixed-order perturbation theory}
\label{nloexpansion}
\renewcommand{\theequation}{D\arabic{equation}}
\setcounter{equation}{0}

Here we evaluate our resummed expressions for the hard matching coefficient $\left|C_V\right|^2$ and the kernels $\bar C_{q\bar q\leftarrow ij}$ in fixed-order perturbation theory, in terms of the coupling $a_s=\alpha_s(\mu)/(4\pi)$. We choose $\mu\sim q_T$, so that $L_M=\ln(M_Z^2/q_T^2)$ is the large logarithm, whereas $L_\mu=\ln(\mu^2/q_T^2)$ counts as an ${\cal O}(1)$ number. 

The one-loop, fixed-order expression for the hard-matching coefficient can be taken from (\ref{CV2fopt}). We find
\begin{equation}
   \left| C_V(-M_Z^2,\mu) \right|^2 
   = 1 + a_s \left[ - \frac{\Gamma_0^F}{2} \left( L_M - L_\mu \right)^2 
    - 2\gamma_0 \left( L_M - L_\mu \right) + C_F \left( \frac{7\pi^2}{3} - 16 \right) \right]
    + {\cal O}(a_s^2) \,.
\end{equation}
There is no need to include higher-order terms in this expression as long as we restrict ourselves to the spectrum at $q_T\ne 0$. To obtain the fixed-order expansion of the kernel functions, we start from our resummed expressions obtained using the $\epsilon$ expansion and reexpand them to second order in $a_s$. This gives
\begin{equation}\label{myresult}
\begin{aligned}
   &\, q_T^2\,C_{q\bar q\to ij}(z_1,z_2,q_T^2,M_Z^2,\mu) \\
   &= \Big[ a_s \left( \Gamma_0^F L_M + 2\gamma_0^q \right) + a_s^2\,k_1(L_M,L_\mu) \Big]\,
    \delta(1-z_1)\,\delta(1-z_2)\,\delta_{qi}\,\delta_{\bar qj} \\
   &\quad\mbox{}+ \left[ \frac{a_s}{2} + a_s^2\,k_2(L_M,L_\mu) \right]
    \left[ {\cal P}_{q\leftarrow i}^{(1)}(z_1)\,\delta(1-z_2)\,\delta_{\bar qj}
    + \delta(1-z_1)\,\delta_{qi}\,{\cal P}_{\bar q\leftarrow j}^{(1)}(z_2) \right] \\
   &\quad\mbox{}+ a_s^2\,\bigg\{ \Gamma_0^F L_M\,  
    \Big[ {\cal R}_{q\leftarrow i}(z_1)\,\delta(1-z_2)\,\delta_{\bar qj}
    + \delta(1-z_1)\,\delta_{qi}\,{\cal R}_{\bar q\leftarrow j}(z_2) \Big] \\
   &\hspace{18mm}\mbox{}- \frac{L_\mu}{4}
    \Big[ {\cal D}_{q\leftarrow i}(z_1)\,\delta(1-z_2)\,\delta_{\bar qj}
    + \delta(1-z_1)\,\delta_{qi}\,{\cal D}_{\bar q\leftarrow j}(z_2) \Big] \\
   &\hspace{18mm}\mbox{}- \frac{L_\mu}{2}\,{\cal P}^{(1)}_{q\leftarrow i}(z_1)\,
    {\cal P}^{(1)}_{\bar q\leftarrow j}(z_2) \bigg\} \,,
\end{aligned}
\end{equation}
where
\begin{eqnarray}\label{k1k2}
   k_1(L_M,L_\mu) 
   &=& L_\mu \left( \Gamma_0^F L_M + 2\gamma_0^q \right) 
     \left[ \Gamma_0^F \left( \frac12\,L_\mu - L_M \right) - 2\gamma_0^q + \beta_0 \right]
    + \Gamma_1^F L_M + 2 \left(\Gamma_0^F\right)^2 \zeta_3 \,, \nonumber\\
   k_2(L_M,L_\mu) 
   &=& - L_\mu \left( \Gamma_0^F L_M + 2\gamma_0^q \right) + \frac{\Gamma_0^F}{4}\,L_\mu^2
    + \frac{\beta_0}{2}\,L_\mu  \,.
\end{eqnarray}
We emphasize the important fact that, due to the collinear factorization anomaly discussed in Section~\ref{sec:fact}, the kernel functions $\bar C_{q\bar q\leftarrow ij}$ contain a dependence on the hard scale $M_Z$ via the large logarithms $L_M=\ln(M_Z^2/q_T^2)$. These large logarithms are not resummed through the evolution of the hard matching coefficient $\left| C_V\right|^2$, but as shown in (\ref{Cdef}) and (\ref{Cfinal}) they exponentiate in $x_T$ space. At $n^{th}$ order in perturbation theory, the Born-level structure $\delta_{qi}\,\delta_{\bar qj}$ in (\ref{myresult}) receives corrections of order $\left(a_s L_M\right)^n$, which are of ${\cal O}(1)$ in RG counting and must be resummed to all orders to accomplish a consistent resummation with NLL accuracy. As we have already mentioned in Section~\ref{sec:fact}, the approach of \cite{Mantry:2009qz} fails to resum the large logarithms in the kernel functions $\bar C_{q\bar q\leftarrow ij}$ (which are written in the form ${\cal G}={\cal I}_n\otimes{\cal I}_{\bar n}\otimes S^{-1}$ in this paper). The one-loop logarithmic term $a_s\,\Gamma_0^F L_M$ in the first structure in (\ref{myresult}) was reproduced in this approach and it resulted from the convolution of collinear beam functions with an inverse soft function. If it could be shown that, to all orders in perturbation theory, only a single large logarithms arises from SCET convolution integrals, then indeed there were no need to resum them and the approach of \cite{Mantry:2009qz} would be consistent. However, already the presence of the two-loop logarithmic term $-a_s^2\left(\Gamma_0^F\right)^2\!L_M^2 L_\mu$ entering via the coefficient $k_1$ in (\ref{k1k2}) shows that this cannot be the case. The presence of this term was missed in \cite{Mantry:2010mk} by making the scale choice $\mu=q_T$, which sets $L_\mu=0$. With any other scale choice a quadratic term $\sim a_s^2 L_M^2$ is required by RG invariance. The peculiar feature that this term vanishes for $\mu=q_T$ is due to the fact that in the $\overline{\rm MS}$ scheme the anomalous exponent for Drell-Yan production does not contain a non-trivial constant at one-loop order, $F_{q\bar q}(L_\perp,a_s)=a_s\,(\Gamma_0^F L_\perp+d_1^q)$ with $d_1^q=0$, see (\ref{Fhexpansions}). In any other scheme except the $\overline{\rm MS}$ scheme, the coefficient $k_1$ contains a term $-d_1^q\,\Gamma_0^F L_M^2$. We also note that in other applications of the collinear anomaly, such as in the study of resummation for the jet broadening distribution in $e^+ e^-$ annihilations \cite{Becher:2011pf}, the corresponding one-loop coefficient $d_1$ is non-zero. 

Adding the contributions from the hard function, we find that the product $q_T^2 \left| C_V(-M_Z^2,\mu)\right|^2$ $\times C_{q\bar q\to ij}(z_1,z_2,q_T^2,M_Z^2,\mu)$ obeys the same decomposition as shown in (\ref{myresult}), with coefficients $k_{1,2}$ replaced by $\hat k_{1,2}$ given by 
\begin{equation}
\begin{aligned}
   \hat k_1(L_M,L_\mu) 
   &= \left( \Gamma_0^F L_M + 2\gamma_0^q \right) 
    \left[ - \frac{\Gamma_0^F}{2}\,L_M^2 - 2\gamma_0^q\,L_M + \beta_0\,L_\mu
    + C_F \left( \frac{7\pi^2}{3} - 16 \right) \right] \\    
   &\quad\mbox{}+ \Gamma_1^F L_M + 2 \left(\Gamma_0^F\right)^2 \zeta_3 \,, \\
   \hat k_2(L_M,L_\mu) 
   &= - \frac{\Gamma_0^F}{4}\,L_M^2 - \gamma_0\,L_M
    + C_F \left( \frac{7\pi^2}{6} - 8 \right)
    - \frac12 \left( \Gamma_0^F L_M + 2\gamma_0^q \right) L_\mu 
    + \frac{\beta_0}{2}\,L_\mu \,.
\end{aligned}
\end{equation}
Note that in this result only single logarithms $L_\mu$ remain. It is a remarkable fact that with our improved expansion scheme based on the modified power counting these scale-dependent logarithms are precisely those ensuring that the differential cross section is scale invariant through ${\cal O}(a_s^2)$. This did not have to be the case, since at two-loop order $L_\mu$ terms not accompanied by a large logarithm $L_M$ are in principle beyond the accuracy of our NNLL calculation. For example, performing the resummation based on a conventional expansion in powers of $a_s$ would miss the last two lines of (\ref{myresult}) as well as some other $\mu$-dependent terms. The fact that with our modified power counting we do account for all scale-dependent terms at ${\cal O}(a_s^2)$ explains why in this case the bands are much narrower than those obtained with a conventional power counting, see the red and green bands in Figure~\ref{fig:3regions}. 

\end{appendix}


\begin{thebibliography}{99}

\bibitem{DDT}
 Yu.~L.~Dokshitzer, D.~I.~Dyakonov and S.~I.~Troyan, 
 Phys.\ Rep.\ {\bf 58}, 269 (1980).

\bibitem{Parisi:1979se}
 G.~Parisi and R.~Petronzio,
 Nucl.\ Phys.\  B {\bf 154}, 427 (1979).

\bibitem{Curci:1979bg}
 G.~Curci, M.~Greco and Y.~Srivastava,
 Nucl.\ Phys.\  B {\bf 159}, 451 (1979).

\bibitem{Collins:1984kg}
 J.~C.~Collins, D.~E.~Soper and G.~F.~Sterman,
 Nucl.\ Phys.\  B {\bf 250}, 199 (1985).

\bibitem{Becher:2010tm}
  T.~Becher and M.~Neubert,
  Eur.\ Phys.\ J.\  C {\bf 71}, 1665 (2011)
  [arXiv:1007.4005 [hep-ph]].

\bibitem{Manohar:2003vb}
  A.~V.~Manohar,
  Phys.\ Rev.\  D {\bf 68}, 114019 (2003)
  [arXiv:hep-ph/0309176].
  
\bibitem{Becher:2006nr}
 T.~Becher and M.~Neubert,
 Phys.\ Rev.\ Lett.\  {\bf 97}, 082001 (2006)
 [arXiv:hep-ph/0605050].
  
\bibitem{Beneke_talk}
M.~Beneke, lectures delivered at the Helmholtz International Summer School on {\em Heavy Quark Physics}, Dubna, Russia, June 2005\\ 
(http://theor.jinr.ru/$\sim$hq2005/Lectures/Beneke/Beneke-Dubna-05.pdf).

\bibitem{Gatheral:1983cz}
 J.~G.~M.~Gatheral,
 Phys.\ Lett.\  B {\bf 133}, 90 (1983).
  
\bibitem{Frenkel:1984pz}
  J.~Frenkel and J.~C.~Taylor,
  Nucl.\ Phys.\  B {\bf 246}, 231 (1984).
  
\bibitem{Chiu:2007dg}
 J.~y.~Chiu, F.~Golf, R.~Kelley and A.~V.~Manohar,
 Phys.\ Rev.\  D {\bf 77}, 053004 (2008)
 [arXiv:0712.0396 [hep-ph]].
 
\bibitem{Becher:2011pf}
  T.~Becher, G.~Bell and M.~Neubert,
  arXiv:1104.4108 [hep-ph], to appear in Phys.\ Lett.~B.
 
\bibitem{Mantry:2009qz}
 S.~Mantry and F.~Petriello,
 Phys.\ Rev.\  D {\bf 81}, 093007 (2010)
 [arXiv:0911.4135 [hep-ph]].

\bibitem{Mantry:2010mk}
  S.~Mantry and F.~Petriello,
  Phys.\ Rev.\ D {\bf 83}, 053007 (2011)
  [arXiv:1007.3773 [hep-ph]].

\bibitem{Becher:2006mr}
 T.~Becher, M.~Neubert and B.~D.~Pecjak,
 JHEP {\bf 0701}, 076 (2007)
 [arXiv:hep-ph/0607228].
 
\bibitem{Ahrens:2008qu}
  V.~Ahrens, T.~Becher, M.~Neubert and L.~L.~Yang,
  Phys.\ Rev.\  D {\bf 79}, 033013 (2009)
  [arXiv:0808.3008 [hep-ph]].

\bibitem{Ahrens:2008nc}
  V.~Ahrens, T.~Becher, M.~Neubert and L.~L.~Yang,
  Eur.\ Phys.\ J.\  C {\bf 62}, 333 (2009)
  [arXiv:0809.4283 [hep-ph]].

\bibitem{Magnea:1990zb}
  L.~Magnea and G.~F.~Sterman,
  Phys.\ Rev.\  D {\bf 42}, 4222 (1990).

\bibitem{Ellis:1981hk}
  R.~K.~Ellis, G.~Martinelli and R.~Petronzio,
  Nucl.\ Phys.\ B {\bf 211}, 106 (1983).

\bibitem{Arnold:1988dp}
  P.~B.~Arnold and M.~H.~Reno,
  Nucl.\ Phys.\ B {\bf 319}, 37 (1989).

\bibitem{Gonsalves:1989ar}
  R.~J.~Gonsalves, J.~Pawlowski and C.-F.~Wai,
  Phys.\ Rev.\ D {\bf 40}, 2245 (1989).
  
\bibitem{Ellis:1997ii}
  R.~K.~Ellis and S.~Veseli,
  Nucl.\ Phys.\  B {\bf 511}, 649 (1998)
  [arXiv:hep-ph/9706526].

\bibitem{Ladinsky:1993zn}
  G.~A.~Ladinsky and C.~P.~Yuan,
  Phys.\ Rev.\  D {\bf 50}, 4239 (1994)
  [arXiv:hep-ph/9311341].
  
\bibitem{Konychev:2005iy}
  A.~V.~Konychev and P.~M.~Nadolsky,
  Phys.\ Lett.\ B {\bf 633}, 710 (2006) 
  [arXiv:hep-ph/0506225].

\bibitem{Martin:2009iq}
  A.~D.~Martin, W.~J.~Stirling, R.~S.~Thorne and G.~Watt,
  Eur.\ Phys.\ J.\ C {\bf 63}, 189 (2009) 
  [arXiv:0901.0002 [hep-ph]].
  
\bibitem{Affolder:1999jh}
  A.~A.~Affolder {\it et al.} [{\sc CDF} Collaboration],
  Phys.\ Rev.\ Lett.\  {\bf 84}, 845 (2000) 
  [arXiv:hep-ex/0001021].
  
\bibitem{Abbott:1999yd}
  B.~Abbott {\it et al.} [D\O\ Collaboration],
  Phys.\ Rev.\ Lett.\  {\bf 84}, 2792 (2000) 
  [arXiv:hep-ex/9909020].
  
\bibitem{Ball:2011mu}
  R.~D.~Ball {\it et al.},
  Nucl.\ Phys.\ B {\bf 849}, 296 (2011) 
  [arXiv:1101.1300 [hep-ph]].
  
\bibitem{vrap} 
L.~J.~Dixon, http://www.slac.stanford.edu/$\sim$lance/Vrap/.

\bibitem{Anastasiou:2003ds}
  C.~Anastasiou, L.~J.~Dixon, K.~Melnikov and F.~Petriello,
  Phys.\ Rev.\ D {\bf 69}, 094008 (2004) 
  [arXiv:hep-ph/0312266].
  
\bibitem{:2007nt}
  V.~M.~Abazov {\it et al.} [D\O\ Collaboration],
  Phys.\ Rev.\ Lett.\  {\bf 100}, 102002 (2008) 
  [arXiv:0712.0803 [hep-ex]].
    
\bibitem{Abazov:2010kn}
  V.~M.~Abazov {\it et al.} [D\O\ Collaboration],
  Phys.\ Lett.\ B {\bf 693}, 522 (2010) 
  [arXiv:1006.0618 [hep-ex]].
  
\bibitem{Aad:2011gj}
  G.~Aad {\it et al.} [ATLAS Collaboration],
    arXiv:1107.2381 [hep-ex].
   
\bibitem{Vesterinen:2008hx}
  M.~Vesterinen and T.~R.~Wyatt,
  Nucl.\ Instrum.\ Meth.\ A {\bf 602}, 432 (2009) 
  [arXiv:0807.4956 [hep-ex]].
  
\bibitem{Banfi:2010cf}
  A.~Banfi, S.~Redford, M.~Vesterinen, P.~Waller and T.~R.~Wyatt,
  Eur.\ Phys.\ J.\ C {\bf 71}, 1600 (2011)
  [arXiv:1009.1580 [hep-ex]].
  
\bibitem{Abazov:2010mk}
  V.~M.~Abazov {\it et al.} [D\O\ Collaboration],
  Phys.\ Rev.\ Lett.\  {\bf 106}, 122001 (2011)
  [arXiv:1010.0262 [hep-ex]].
 
\bibitem{Banfi:2011dx}
  A.~Banfi, M.~Dasgupta and S.~Marzani,
  Phys.\ Lett.\ B {\bf 701}, 75 (2011)
  [arXiv:1102.3594 [hep-ph]]. 

\bibitem{Marzani:2011yf}
  S.~Marzani, A.~Banfi, M.~Dasgupta and L.~Tomlinson,
   arXiv:1106.6294 [hep-ph].
 
\bibitem{qt}
R.~Gonsalves, http://www.physics.buffalo.edu/gonsalves/.
 
\bibitem{Bozzi:2010xn}
  G.~Bozzi, S.~Catani, G.~Ferrera, D.~de Florian and M.~Grazzini,
  Phys.\ Lett.\  B {\bf 696}, 207 (2011)
  [arXiv:1007.2351 [hep-ph]].
  
\bibitem{Balazs:1997xd}
  C.~Balazs and C.~P.~Yuan,
  Phys.\ Rev.\ D {\bf 56}, 5558 (1997) 
  [arXiv:hep-ph/9704258].

\bibitem{Landry:2002ix}
  F.~Landry, R.~Brock, P.~M.~Nadolsky and C.~P.~Yuan,
  Phys.\ Rev.\ D {\bf 67}, 073016 (2003) 
  [arXiv:hep-ph/0212159].

\bibitem{Mantry:2011xj}
  S.~Mantry and F.~Petriello,
    arXiv:1108.3609 [hep-ph].
        
\bibitem{Neubert:2004dd}
  M.~Neubert,
  Eur.\ Phys.\ J.\  C {\bf 40}, 165 (2005)
  [arXiv:hep-ph/0408179].
    
\end{thebibliography}
\end{document}